# ESO Future of Multi-Object Spectroscopy Working Group Report

06-09-2016

ESO Future of Multi-Object Spectroscopy Working Group

Richard S Ellis (ESO), Joss Bland-Hawthorn (Sydney), Malcolm Bremer (Bristol), Jarle Brinchmann (Leiden), Luigi Guzzo (Milan), Johan Richard (Lyon), Hans-Walter Rix (Heidelberg), Eline Tolstoy (Groningen), Darach Watson (Copenhagen)

## Executive Summary

We consider the scientific case for a large aperture (10-12m class) optical spectroscopic survey telescope with a field of view comparable to that of LSST. We find that such a facility could enable transformational progress in several broad areas of astrophysics, and may constitute an unmatched ESO capability for decades. Deep imaging from LSST and Euclid will provide accurate photometry for spectroscopic targets beyond the reach of 4m class instruments. We discuss the scientific potential of such a facility in undertaking ambitious new surveys in Galactic and extragalactic astronomy. It would revolutionise our understanding of the assembly and enrichment history of the Milky Way and the role of dark matter through chemo-dynamical studies of tens of millions of stars in the Local Group. Extending `chemical tagging', whose effectiveness scales as the square of the sample size, and including a wider list of elements such as those beyond the iron peak, will lead to fundamental discoveries. Emission and absorption line spectroscopy of redshift z~2-5 galaxies can be used to directly chart the evolution of the `cosmic web' and examine its connection with activity in galaxies. The facility will also have synergistic impact, e.g. in following up `live' and `transpired' transients found with LSST, as well as providing targets and the local environmental conditions for follow-up studies with E-ELT and future space missions. For each scientific case, we present the requirements for the telescope and its spectrographs. Although our study is exploratory, we highlight a specific telescope design with a 5 square degree field of view and an additional focus that could host a next-generation panoramic IFU. We discuss some technical challenges and operational models and recommend a conceptual design study aimed at completing a more rigorous, broadly-based science case in the context of a costed technical design. ESO can take the lead in defining an area ripe for exciting science in the next decade and, noting the financial challenges, may wish to consider establishing links with other international communities given the evident interest in having such a facility in the southern hemisphere.

# 1. Scope of the Working Group and Methodology

The Working Group (WG) was established by the ESO Director for Science in response to the ESO Scientific Strategy WG, discussions at the workshop "ESO in the 2020s" held in January 2015 which followed the results of a community poll[1] conducted in late 2014. That poll revealed two significant findings of relevance to the present report. In response to the question *What is the most important capability for your research in 2020-2030?*, the largest number of responses refer to Wide field spectroscopic surveys and a high multiplex gain with high spectral resolution. With respect to wavelength, the greatest demand was for optical (0.4-1 microns) and near-infrared (1-2.4 microns). Addressing the question *What facilities are most required for your research in 2020-2030?*, a dedicated optical/infrared spectroscopic 10-metre class telescope ranked fourth but was the highest requested facility not yet in current use nor under construction.

The charge for the WG was agreed between the Director for Science and the WG chair and is listed in Appendix A-1 together with membership details. Our goal has been to examine the scientific potential for a future wide-field spectroscopic telescope in the context of projected ground and space-based imaging surveys, as well as extant or projected multi-object spectrographs being developed for 4 and 8-meter telescopes and the capabilities of anticipated instrumentation on the E-ELT. Given multi-object spectroscopy now includes panoramic integral field systems such as MUSE which can conduct spectroscopic surveys without relying on pre-determined photometric targets, the WG was also asked to examine the future role of such integral field instruments.

The WG rationale has been to explore the merit of a few ambitious science questions that will be driving astronomical research in the 2020s, demonstrating that these are beyond the capabilities of upcoming facilities. Additional considerations include synergies with complementary facilities, the most obvious of which include LSST, the panoramic imaging telescope, and the E-ELT, which will be the most powerful of its generation but with a relative small field of view. Ultimately the case for a new survey facility lies with its improved capability which could be in terms of survey depth, spectral resolution and/or survey size. We stress that our strategy at this early stage has been to consider a limited number of programmes that define a broad set of technical requirements. Although we have examined some of the relevant technical issues, further work would be necessary before a selection of telescope and instrument designs can be presented. Details of the WG meetings are summarised in Appendix A-1.

The WG has taken full advantage of other international study reports relating to wide-field large aperture spectroscopic survey telescopes (see Appendix A-2 for a complete summary of consulted documentation). For one of these, the Mauna Kea Spectroscopic Explorer (MSE), there is a very detailed justification for a specific facility.

---

[1] ESO Messenger **161**, September 2015

As emphasized above, the goal of this report is not to propose a specific facility in any technical detail, but rather to establish that a dedicated spectroscopic survey capability exploiting a 10-12 metre aperture telescope is an exciting one scientifically notwithstanding the current investment in 4-8 metre multi-fibre spectrographs. We make specific recommendations on the next steps to further justify the scientific promise of such a capability in terms of cost and technical feasibility in Section 8.

## 2. Importance of Survey Spectroscopy

Spectroscopy will always be a primary tool of ground-based astronomy yielding unique astrophysical insight into the chemical composition and radial velocities of stars in the Milky Way and nearby resolved galaxies, and accurate redshifts, measures of internal motions, the nature of stellar populations, non-thermal sources and the ionizing radiation field in a variety of extragalactic sources over cosmic time.

The last decade has seen a huge investment in survey imaging at optical and near-infrared wavelengths. At ESO this is manifest in surveys conducted with the VST and VISTA telescopes. Elsewhere, common-user facilities have been commandeered to undertake large surveys such as the Dark Energy Survey (DES) at the Blanco telescope at Cerro Tololo and that being undertaken with the Hyper Suprime-Cam panoramic imager on the Subaru 8.2m telescope. Even more ambitious surveys are planned with the Large Synoptic Survey Telescope (LSST) now under construction, and in space with the ESA Euclid and NASA WFIRST missions. Collectively, these imaging facilities will provide new catalogues of spectroscopic targets with improved reliability and to much fainter limits than is presently possible. This paves the way for a spectroscopic facility that can exploit this rich photometric data in numerous ways as well as provide the key diagnostic and local environmental data for targets observed with facilities at other wavelengths e.g. ALMA and the Athena X-ray mission.

The Sloan Digital Sky Survey (SDSS) and surveys with the UK-Australian 2 degree field (2dF) instrument have clearly shown the benefit of combining imaging surveys with matched spectroscopic data. In the case of SDSS, the telescope was equipped with both an imaging and a multi-fibre spectroscopic capability. In the case of 2dF, spectroscopic targets were pre-selected from all-sky photographic surveys. Remarkably, 15 years on, these telescopes continue to exploit their basic infrastructure to do new surveys.

The AAT 2dF facility has undertaken new surveys such as WiggleZ (a baryonic acoustic oscillation survey to intermediate redshift), GAMA (an extension of the original 2dF galaxy survey to include lower luminosity galaxies), HERMES-GALAH (a high spectral resolution Galactic survey) and SAMI (a multi-IFU survey). Likewise the SDSS infrastructure has enabled the SEGUE-1/2 Galactic surveys, BOSS and e-BOSS (large scale structure surveys), APOGEE-1/2 (stellar surveys at near-infrared wavelengths) and MaNGA (a multi-IFU survey).

Two important lessons can be learned from the SDSS and 2dF history. Firstly, *a powerful spectroscopic survey facility is likely to remain productive for several decades*. Modest reconfigurations with updated detectors and/or new instrumentation can address new and presently unforeseen scientific questions. This will particularly be the case if the facility is designed to be versatile, for example with accessible focal planes. Secondly, *large unique datasets yield scientific surprises!* These could include the spectroscopic identification of new transient phenomena first located by LSST, or fundamental progress in stellar physics from unforeseen trends located in neutron rich elements.

## 3. Current and Future Spectroscopic Facilities

Given the explosion in imaging surveys discussed above, it is not surprising that there is now significant interest in associated spectroscopic follow-up of the targets being delivered and therefore already a considerable investment in new multi-object spectroscopic facilities. It is convenient to consider them in three broad categories (see also Appendix A-3 for a selected summary and basic parameters)

1. The conversion or upgrading of existing 4m class telescopes for long-term multi-fiber optical surveys following the "2dF model". These include 4MOST on the VISTA telescope, WEAVE at the 4.2m William Herschel Telescope and the Dark Energy Spectroscopic Instrument (DESI) at the 4.0m Kitt Peak Mayall telescope. With regard to 2dF, these provide new capabilities, for example higher spectral resolutions (4MOST, WEAVE) and/or a larger field of view and multiplex gain (DESI). Ultimately, however, they will be limited in performance by the same 4 metre platform. Experience has shown that longer exposures on dedicated 4m facilities never match the increased scientific parameter space possible with 8-10m facilities.

2. Multi-fiber instruments on common-user 8m telescopes exploiting a larger aperture and providing extensions in the near-infrared spectral region. MOONS has just passed its PDR for the ESO VLT and offers a 1000 fibre system within a 0.14 deg$^2$ Nasmyth field of view spanning the red and near-infrared spectral region (0.64 – 1.63 microns). The Subaru Prime Focus Spectrograph (PFS) offers 2400 fibres within a larger 1.25 deg$^2$ field; it has a more restricted range of spectral resolutions but spans a wider wavelength range from 0.38 – 1.30 microns. In this category, but unfunded at present, is the Mauna Kea Spectroscopic Explorer (MSE) which is a proposed 11.25m telescope designed to replace the CFHT 3.6m telescope. It is likely to have a 1.5 deg$^2$ field and several thousand fibres. Only MSE represents a dedicated survey telescope but its future is currently uncertain given the moratorium for construction of new facilities in Hawaii.

3. Panoramic integral field units (IFUs) on 8-10m telescopes offering the capability of undertaking `blind' spectroscopic searches without reliance on photometrically defined targets. These include MUSE on the VLT which offers optical spectroscopy over a maximum field of 1 arcmin$^2$, the Keck Cosmic Web Imager (KCWI) with similar characteristics, and HETDEX, an array of 150 deployable IFUs on the zenith pointing Hobby-Eberly 10m telescope.

There are many aspects to consider in judging whether there is a case for a further dedicated spectroscopic facility. What science will these upcoming facilities undertake and what questions will they be unable to undertake, either due to their limited aperture (category 1 above), their restricted survey speed via their limited field of view (e.g. MOONS), or the fact they share their telescope with other instruments (PFS, MOONS)? Currently MUSE is in great demand as the community appreciates its new capabilities. But is there a case for an even more powerful MUSE-like capability?

## 4. Science Perspectives

Experience has shown that most astronomical facilities scientifically outperform their original justifications. This is in part because of the imagination and determination of astronomers who push each facility beyond its original goal, but largely because our subject unfolds in ways that weren't predicted. Defining a detailed science case for a facility at least a decade in the future is challenging, even more so because several relevant multi-object spectroscopic facilities have yet to gather data.

In evaluating the present case, the WG considered 4MOST to be a valuable benchmark but MOONS and PFS to be the most potentially far-reaching upcoming facilities noting that MSE, although well-conceived, remains unfunded. Although MOONS and PFS share their telescopes with other instruments and are currently slated to undertake specific surveys (e.g. PFS is conceived to undertake a 300 night Strategic Survey Programme over 5 years), it is reasonable to assume they will eventually become semi-dedicated facilities in the late 2020s. MSE is a well-defined dedicated project with a broad and comprehensive science case and its documentation was valuable input to the WG.

Although we later make specific recommendations on the steps that could be taken to define a more inclusive science case for a future facility, at this stage our WG focused on 3 themes which will complement the likely achievements of PFS and MOONS. In each case we examine the merits of a greater étendue (the product of field of view, aperture and multiplex gain) and spectroscopic resolution. Where relevant, we also explore the potential science arising from a large format integral field spectrograph (referred to here as `Super-MUSE').

(i) An ambitious Galactic and Local Group survey of stars undertaken at high spectral resolution to considerably extend the chemical tagging programmes being considered by upcoming facilities.

(ii) A high signal to noise survey of intermediate redshift ($1 < z < 5$) galaxies capable of charting their 3D distribution across unprecedented cosmic volumes, as well as revealing absorption line signatures arising from their interstellar gas, stars and foreground intergalactic gas clouds, the latter of which can be used to reconstruct the cosmic web of dark matter.

(iii) Science arising from the rapidly-developing area of transient phenomena revealed, for example, by LSST imaging, and more generally studies of other rare sources where the science case is compelling but the surface density of targets may be insufficient to match the requirements determined by the programmes discussed in (i) and (ii).

Noting that these 3 cases are illustrative, rather than a comprehensive science case that would emerge from a more exhaustive community survey, we describe each in turn below and then study the technical requirements that follow from these examples in Section 5.

## 4.1 Galactic Archaeology: The Milky Way as a Model Galaxy Organism

It has long been recognized that our Milky Way and its immediate environment offers a unique opportunity to understand *how disc galaxies work*. The Milky Way is the only large galaxy whose history can be studied using the full distribution of stars from white dwarfs to supergiants. Nearby systems in the extended halo of the Milky Way and the Local Group permit studies of the relation between a large spiral and its environment. This is a critical link for a proper understanding of how galaxies build up their mass, and how and when most stars are formed throughout time.

While most stars reside in systems that are supported by rotation, a substantial fraction reside in systems supported by internal random motions (pressure); these two extremes clearly have very different origins. Moreover, half of all stars in the Galaxy and the majority in many nearby dwarf galaxies were formed before a redshift of unity, enabling us to probe events early in the Universe.

Today we clearly recognize that the stars in the Milky Way and its satellites encode decisive information about galaxy formation in general and the particulars of own Galactic formation history (Freeman & Bland-Hawthorn 2002). The questions that can be addressed by knowing the ages, masses, element composition and orbits of the stars in the Milky Way system, fall into four broad categories:

- *The Galactic gravitational potential and the properties of dark matter*
    - What can we learn about the 3D distribution of dark matter in the Milky Way and its visible satellites? Can we demonstrate direct evidence for the existence of very low-mass (star-less) dark matter halos, one of the most fundamental untested predictions of CDM?

- *The formation history and memory of a prototypical large galaxy*
    - What was the dynamical history of the Galaxy and how does it fit in the overall context of hierarchical cosmology? A comprehensive census of the halo stars in orbit and abundance space is key both for the 3D mass distribution and the merging history.
    - How much (orbital) dynamical memory does a disk galaxy retain? 'Chemical tagging', recognizing widely dispersed stars once born together plays a central role.
    - The bulge of the Milky Way is its most complex and potentially informative part, containing the oldest and most metal rich stars. Yet, the relation between the bulge and the other components are still not understood.

- *Stellar physics and the origin of the chemical elements*
    - How does the interplay between nucleosynthetic yield, star-formation history, stellar mass function, and gas in/outflow lead to the observed intricate [X/H]- abundance pattern that we see to vary with age and orbit within the Galaxy and its satellites.
    - Large samples of detailed measurements are required to fulfill the needs of stellar modellers, e.g. to understand specific questions such as the r-process.
    - Binary and variable stars can be studied with multi-epoch spectra critical for understanding e.g. which abundance patterns can be attributed to binary star evolution.

- *Satellite galaxies as model organisms of low-mass galaxies*
    - How do the nearby small companions of the Milky Way fit into the context of hierarchical cosmology? Nearby dwarfs are valuable probes of galaxy formation at lower mass and lower [Fe/H]. The Magellanic Clouds are key to understanding the impact of merging on the stars and gas in fully-formed galaxies on a large scale.

These goals require large, wide-field and most of all *accurate* datasets, providing precision velocities, proper motions and abundances of a large range of elements, especially those beyond the Iron-peak where greatest variety in patterns across different stellar systems is seen.

**Upcoming Stellar Spectral Surveys and Beyond**

There are ~85 million stars brighter than V < 17 across the South-accessible sky (see Figure 1) and ~30 million with V < 15.5. All are, in principle, amenable for determining the ultimate set of observables, which is the joint distribution of kinematics, physical properties and abundances, viz. $n(r, v,$ log g, $T_{eff}$, $M_*$, $v_{turb}$, .. [Fe/H], [X/H], $t_{age}$). The ultimate goal is to have these measures for *all* stars in the Galaxy and its satellites for which these can be measured, providing a fully complete overview of the star formation and chemical evolution history of the Milky Way and its nearest satellites. The attributes of each star fall into three groups: a) its orbit or kinematic properties ($r$, $v$ and $\Phi(r)$), b) its stellar (evolutionary) parameters, and c) a range of elemental abundances, reflecting (most often) the composition of the star's birth material. Finally, the star's age, a notoriously difficult parameter to determine, relates to all these aspects.

The Gaia mission and accompanying spectroscopic surveys will revolutionize our knowledge of ($r$, $v$), provide $T_{eff}$, log g, [Fe/H], [α/Fe] and ages for a large subset of mostly nearby stars. Astroseismology, is also starting to provide extremely precise stellar parameters for selected stars in the Galaxy, and this number will continue to increase in the future. Ongoing and planned spectral surveys of stars using the upcoming facilities (discussed in Section 3) will provide an excellent basis to efficiently measure a broader range of abundances in stars for which the basic atmospheric parameters will, by then, be well established.

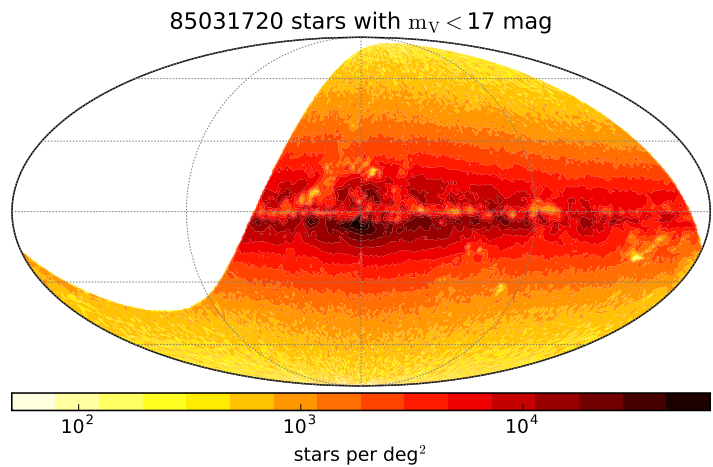

*Figure 1*: Distribution of the 85 million stars brighter than V < 17 across the Southern sky, based on the Galaxia model; courtesy J. Rybitzki.

By 2025, we will be in the post-Gaia era and the LSST 10-year survey will be underway providing multiband photometry 4 magnitudes deeper than SDSS. Gaia will provide proper motions to V~20 and accurate radial velocities (<1 km s$^{-1}$) for ~200M bright V < 14 stars. The southern GAIA-ESO and GALAH surveys will secure velocities to a fainter limit of V~16 and be followed by MOONS and 4MOST greatly expanding the area covered to a similar depth. A new facility with a wider field of view and greater sensitivity would take the subject forward dramatically, probing precision velocities to V~20, thereby fully exploiting Gaia's proper motions for studying kinematic substructures (e.g. cold streams) in all components of the Galaxy (Bovy et al 2015). Most importantly, it would provide high resolution spectra to V~17 ensuring precision abundances of numerous elements as discussed below.

To obtain detailed chemical abundances for a wider range of elements, *higher quality and higher spectral resolution* spectra from a next generation facility are needed for the following reasons:

- There is rich and complex structure in the distribution of chemical elements in the stellar populations of the Milky Way and its satellites. To disentangle the implications requires precise measurement of multiple chemical elements created by different nucleosynthetic channels (i.e. at least 20-30 elements which can be grouped into several independent sets (e.g. α-process, r-process etc.).
- Accurate abundances from stellar spectra with a dense and spatially contiguous coverage of a large fraction of the full sky are needed to recognize all the large-scale features in orbit-abundance space (e.g. stellar streams, warps, ultra faint dwarfs) in both the halo and disc of the Milky Way.
- Huge wide-field samples are also needed to find rare objects, whose abundances are exceptionally informative about chemical enrichment processes (e.g. about SN yields or other enrichment and dispersion processes in a wide range of environments).
- Current/planned large area (≥ 10.000 deg$^2$) surveys mostly target stars within a few kpc of the Sun, and because of limitations in wavelength coverage or spectral resolution are only able to provide a limited view of the landscape of abundances necessary for detailed chemical tagging.

In summary, the current and planned spectroscopic surveys discussed in Section 3 will require more detailed follow-up to provide the ultimate mapping of the kinematics and abundances for a major fraction of the Milky Way population. They will set the scene but only provide spectra for ≤ 1% of the stars in our quadrant of the Milky Way, the halo and some satellite dwarf galaxies. Generally, they will provide a limited number of element abundances with precision of ≤ 0.1 dex over a restricted volume. Ideally we seek to probe a range of distance scales - 8 kpc for the disc, 20 kpc for the halo, and for the satellites from ~ 20−130 kpc and beyond.

**Importance of the Southern Hemisphere**

The opportunities for stellar spectroscopic surveys in the context of the Milky Way system are not hemisphere-symmetric: the vast majority of Milky Way stars ($\sim 80\%$) are in the Southern hemisphere. In particular, from an evolutionary perspective, the richest part of the Milky Way is the central few kpc, more accessible in the southern hemisphere. Similarly, owing to the LMC and SMC, the vast majority of stars in Milky Way satellites are also in the south. So far, the majority of large-scale stellar surveys have been undertaken in the northern hemisphere (via LAMOST, APOGEE and SDSS/Segue).

**Detailed Abundances**

Stellar abundances represent a key route to understanding the history and structure of the Milky Way. Even today, new stellar components of the Galaxy are being discovered through their unique chemical signatures, e.g. the APOGEE survey has recently uncovered a nitrogen-rich population of stars uniformly distributed in bulge which may represent disrupted globular clusters (Schiavon et al 2016). Gathering detailed abundances of *many* chemical elements in vast samples of stars with known distances and velocities will clearly represent the frontier beyond 2025 and this exciting prospect is motivated by several important considerations.

Firstly, chemical tagging, the identification of peaks in abundance space is only feasible in a high-n dimensional space (Wheeler et al 1989, Sneden et al 1996, Bland-Hawthorn et al 2010, Recio-Blanco et al 2014, Hogg et al 2016) and with a precision of < 0.05 dex (Ting et al 2015, Figure 2).

Secondly, detailed abundances are key to understanding both stellar physics and galaxy evolution. Different supernova (SN) yields result in markedly different abundance patterns (Woosley & Weaver 1995). These patterns will be most distinct among the oldest stars often thought to have the lowest metallicities given fewer SN events will have contributed to their chemical enrichment. To understand which different types of SNe, or other mass loss mechanisms may be responsible for the early chemical enrichment of the Milky Way and its environment, studying neutron capture elements is essential (McWilliam 1997, Sneden et al 2008). Their study can constrain the physical conditions of star formation at the earliest times as it depends, through the 'mix' of nucleosynthetic channels, on the star formation history, ISM matter cycle and the mass function (Figure 3).

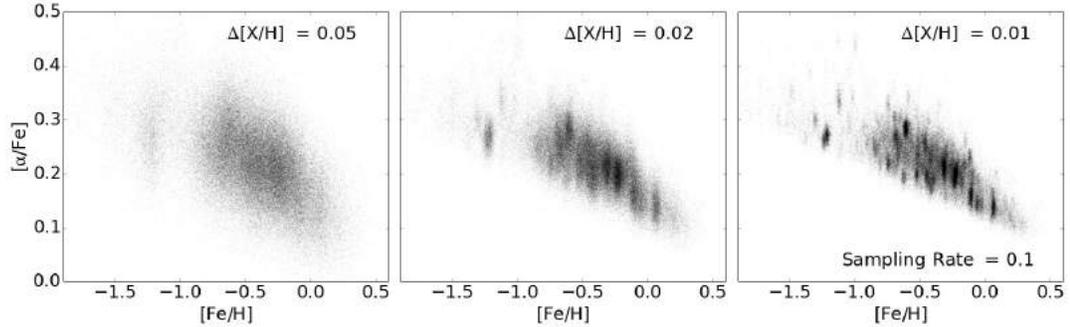

*Figure 2*: *The importance of precise elemental abundances for "chemical tagging". With poor precision, most of the substructures in abundance space is smeared out, as shown in the left panel. If the full information encoded in blended spectral lines can be exploited, higher precision can be achieved, and so many more individual clusters can be identified as shown in the middle and right panels. [Courtesy: Y-S Ting]*

**Beyond the Milky Way**

The Milky Way is a fairly representative large spiral and thus provides the opportunity for a detailed insight into how most of the (stellar) mass in the Universe evolves and changes over time. Dwarf galaxies in the vicinity of the Milky Way indicate how small scale structures evolve. Although they contain a small fraction of the overall stellar mass density, they are much more numerous, and represent important `building blocks' in the hierarchical assembly picture. Many stopped forming stars early so they also provide an unobscured view back to star formation at primaeval times. Their small size makes them extremely sensitive to trace individual enrichment events and offers the best opportunity for identifying chemical enrichment sites.

Stars form with different abundance signatures in different environments. If the building blocks of the Milky Way were similar to those that remain today around it, they must have merged within the first billion years of star formation. Since the "knee" in the [Ca/Fe] vs. [Fe/H] diagram (Figure 3a) falls in different places for different mass galaxies, even if the early star formation processes were similar they must have later diverged. Figure 3b presents the difference in s-process elements between the Milky Way and local dwarfs, illustrating the strong influence of AGB stars at a time when massive stars no longer drive metallicity evolution. Understanding these variations will require much larger samples and more detailed abundance surveys. As the Milky Way satellites are distant this kind of study is beyond the upcoming 4 metre class surveys. Figure 3 shows what is possible with VLT/FLAMES but for only the inner portion of the Sculptor dSph (at R$\sim$ 20,000), a system whose full extent at 86 kpc distance is >2 degrees.

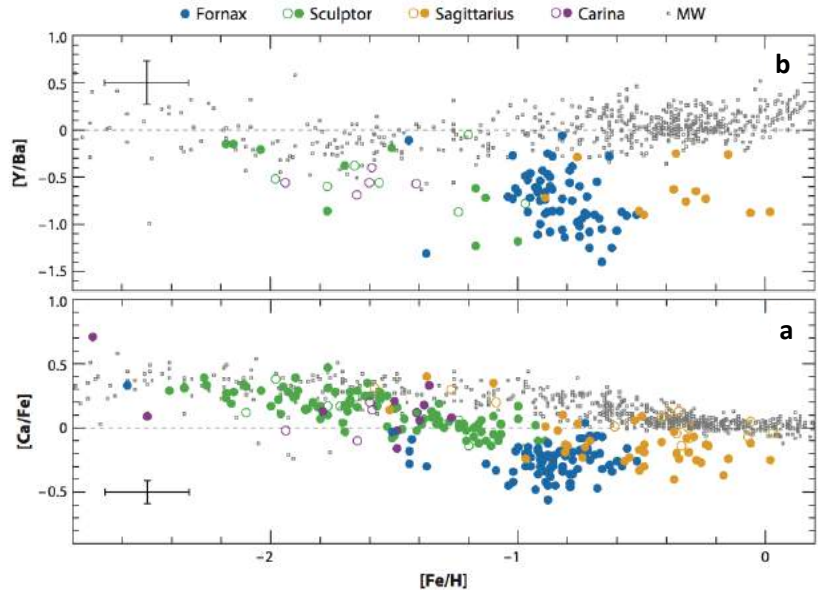

*Figure 3: Observations of individual stars in nearby dwarf galaxies (coloured symbols) compared to the Milky Way (black symbols), illustrating the similarities and differences in abundances of (a) alpha elements [Ca/Fe] and (b) first-to-second s-process peak elements [Ba/Y] and (b) (from Tolstoy et al. 2009).*

Such comprehensive surveys are also of great interest for the Small and Large Magellanic Clouds (50-60kpc distant). They will benefit from the increased sensitivity and higher spectral resolution opportunities in the same way as the Milky Way. These have so far been carried out on a relatively small scale with FLAMES (e.g., Van der Swaelmen et al 2013) and will be the target of more extensive 4MOST surveys. The LMC contains a bar, a disrupted disc and an extended faint stellar halo and the SMC, although more irregular, has also been disrupted. These are the only examples of a late interaction between two gas-rich star-forming galaxies than can be observed in great detail. The effects of merging on the various components, via chemical tagging and kinematics, will be very instructive. Spectroscopic follow up of the Magellanic Bridge and Stream would confirm the age distribution and composition of the stripped material. The extended size of these features on the sky demands a very wide field capability.

The challenge in considering applications for a panoramic integral field spectrograph (the `Super-MUSE') is the necessity for both high angular *and* spectral resolution. There are crowded regions in both the Milky Way and nearby galaxies and a panoramic IFU has an obvious advantage in foregoing the need for a target list. The current VLT MUSE will soon be equipped with low order adaptive optics but its spectral resolution is inadequate for precision abundance studies. Studies of the central regions of the LMC, including the bar and 30

Doradus could determine the impact of the merging process on the detailed star formation properties of this (relatively) low metallicity system ($Z \sim 0.3Z_\odot$). Likewise a panoramic IFU could also probe the centres of nearby dwarf irregular galaxies (e.g. NGC 6822, IC 1613, IC 10) and smaller systems such as Phoenix, Leo T, Cetus & Tucana. These galaxies are valuable non-interacting systems for comparison to the Magellanic Clouds and dSph galaxies found closer to the Milky Way. Detailed abundances and stellar kinematics in star clusters in nearby galaxies also requires an IFU with fairly high spatial & velocity resolution. Some progress is being made with MUSE but this will not be with high spectral resolution.

**Resolution and S/N Requirements**

The accurate determination of stellar parameters and detailed elemental and isotopic abundances for a wide range of elements, encoded in (often blended) photospheric absorption lines, requires spectra with high spectral resolution $R$ and high S/N covering an extensive wavelength range. This is typically only possibly at the present time with single slit spectrographs (e.g., UVES). High spectral resolution (for a given multiplexing) always implies less spectral coverage, and hence fewer lines. The minimum useful S/N and $R$ (and wavelength range) depends on the type of star and thus the number and strength of the useful absorption lines. The abundances are usually compared to detailed models of stellar and chemical evolution, which can make precise predictions. Differentiating between intrinsic scatter in the abundances and measurement errors is a crucial issue. FGK stars with Teff $\sim 4000 - 6000$ K are most suited, and fortunately these are very common and relatively luminous.

It is generally agreed that accurate chemical tagging requires detailed element abundances of a wider range of elements with a precision better than ≤ 0.05 dex. This requires a spectral resolution $R \sim 20,000$–$40,000$ with a S/N>80 as many elements have lines that are very weak (equivalent widths only a few mÅ, corresponding to $\sim$ 5% continuum depth). The resolution should ideally be matched to the intrinsic line width and the elements to be measured (e.g. Hansen et al 2015).

Obtaining accurate abundance measurements for the important r-process element Eu in metal poor stars (Figure 4) ideally requires R$\sim$ 45,000 – 60,000 and S/N>100. The effect of spectral resolution on the ability to identify the Eu line can clearly be seen. There are some elements for which upper limits on their abundances have important implications for chemical evolution. For example, the absence of Zn is a prediction of some zero metallicity stellar evolution models. Thus only measuring certain elements, like Zn and Eu, when they are over-abundant can lead to an incorrect picture of the chemical evolution of a stellar population. If all lines, including blended ones, are included in the simultaneous fitting of all abundances, the full information content of the spectra may enable estimates of similar precision at lower S/N (e.g. Ting et al 2016; Rix et al 2016); but the combined S/N and resolution requirements are currently still under active debate and considerable community effort (e.g. Smiljanic et al 2014, Worley et al 2016).

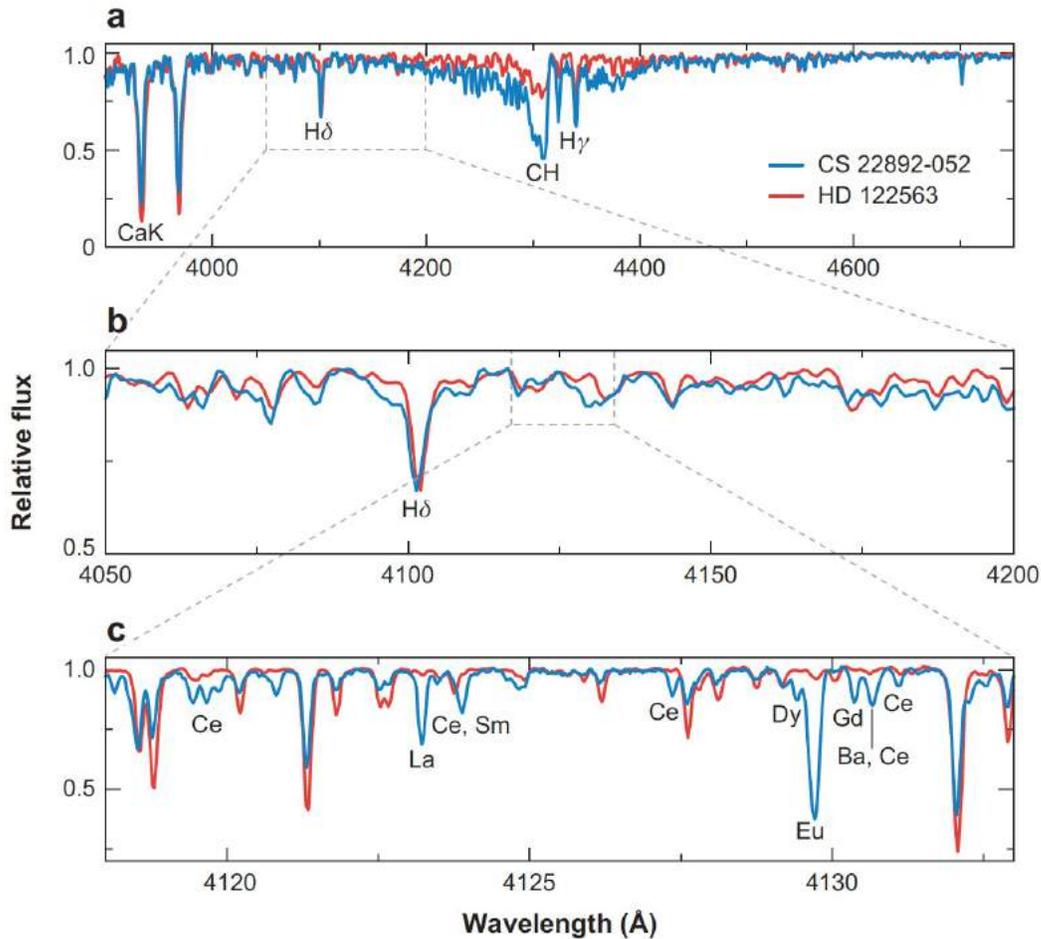

*Figure 4: The need for high resolution spectra to determine accurate abundances of r-process elements, such as Eu, from very weak spectral lines. The red and blue spectra in the upper two panels have R~2500 compared with R~45,000 and ~65,000 on the lowest panel. Many diagnostic lines have depths of only 5-10%, driving the resolution and S/N requirements (Sneden et al 2008).*

**A Proposed Survey**

*Large Survey with Dense Sampling*

The Milky Way is the ultimate $4\pi$ feature in astronomy. As we view it from an off-center vantage point, large scale asymmetries matter as much as the small scale structure. Existing high-resolution spectral surveys have coped with this by

`judiciously sub-sampling' the various components of the Milky Way system, including its satellites.

This has three serious drawbacks:

- The Milky Way is lacking sufficient symmetry for short-cuts: stellar streams in the halo (that may make up much of the halo) are individual features, which therefore have grossly asymmetrical sky coverage; because of the spiral arms and the Galactic bar seen in perspective there are strong asymmetries in the disk plane.
- Many of the Milky Way's features, where the correlation between the orbit and detailed abundances matter, are narrow and contiguous. The contiguous nature of features is all but impossible to recognize for sparse (angular) sampling.
- Most importantly, all variants of "chemical tagging" require dense sampling. For example, the number of stellar siblings one will find, grows quadratically with fraction of all stars sampled, and hence with sample size. While with many other observational experiments the signal grows as the square root of the observational effort, here it grows quadratically. Surveys for high-quality high-resolution spectra of stars will (in sum) provide $\sim$ 3 million spectra of the above quality. Going to 50 million should increase e.g. **300-fold** the number of chemical- tagged twins.

*Access to the blue*

There is no single (sensibly finite) wavelength range that can satisfy all desiderata of Galactic stellar spectroscopy. On the one hand, the majority of the Milky Way's stellar disc is highly dust-obscured, which argues for IR spectroscopy as implemented by APOGEE and MOONS. Yet, atomic physics pushes towards the shortest wavelengths accessible from the ground. Figure 5 illustrates this point: while light elements, $\alpha$-elements and even iron-peak elements have absorption lines across the optical spectrum, the neutron capture elements (*r*-process and *s*-process) that are crucial for chemical tagging and understanding the still-open origin of heavy elements, are mostly concentrated to the blue part of the optical spectrum ($3600 < \lambda < 4500$Å ). This argues for a blue spectroscopic capability, but restricted to stars that are only modestly reddened.

*Go faint*

The Milky Way's stellar halo extends to at least 100 kpc, and most of the Milky Way's satellite galaxies lie well beyond 20 kpc. Understanding these parts of the Milky Way systems requires spectra of $R \geq 20{,}000$, and S/N$\geq$ 50. For a characteristic distance of $\sim$ 50 kpc, this translates into reaching $V \sim 17$. The magnitude range of $V \sim 16 \pm 1$mag is also the regime where Gaia can provide interesting proper motions constraints.

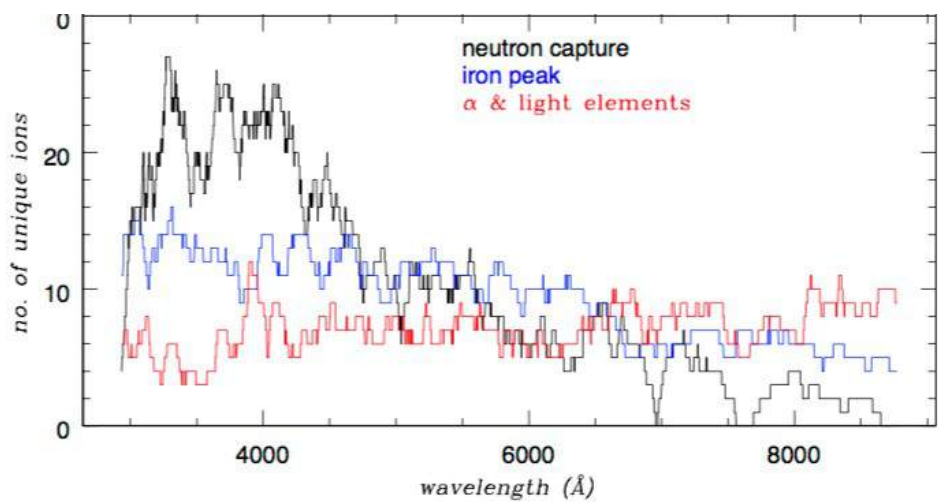

*Figure 5:* Number of stellar absorption lines across the optical spectrum with respect to various groups of elements (Bland-Hawthorn & Freeman 2004). For the neutron capture process diagnostic species, access to the blue spectral region is critical.

*Summary and Requirements*

The primary proposed survey is to secure $R \sim 20{,}000 - 40{,}000$ spectra in selected wavelength regions from 3600 – 9500 Å at a S/N~80 for 30-85M accessible stars to a magnitude limit of V~15.5 – 17 to secure a rich array of chemical and kinematic data across the southern Milky Way (see Figure 1). Assuming a multiplex gain of 5000 and a 10-12m class telescope, exposure times of 1-2 hours should be sufficient ensuring such an ambitious survey would be completed in 5-10 years. As the target density at V<17 ranges from 600-10,000 $\text{deg}^{-2}$, a field of view of ~5 $\text{deg}^2$ is needed to efficiently exploit this multiplex gain. It will be necessary to trade off wide field full Galactic surveys at R~20,000 with more detailed studies of substantial subsets at R ≥ 40,000, taking into account the wavelength ranges these entail. Of course, it may be possible to conduct such studies simultaneously with such a high multiplex gain.

To obtain useful spectra of individual stars in nearby dwarf galaxies, whether the diffuse but closer ultra-faint dwarfs or the larger, more compact and dense systems requires the ability to measure accurate abundances of stars to $V \geq 17.5$ and fainter. Although ultra-faint dwarfs are closer, they contain very few red giant branch stars, and so it is necessary to reach the Main Sequence Turnoff region. Dwarf spheroidals have sufficient numbers of red giants, but are more distant (> 80 kpc). In both cases the target area exceeds most current telescopic fields of view. Longer integrations at the positions of these nearby galaxies and stellar streams could be included as part of an all-sky survey. $R \sim 40{,}000$ spectra with a S/N~30 at $V \sim 19 - 20$ would take ~5 hours.

## 4.2 Extragalactic Science: Galaxy Assembly & the Cosmic Web

**The context**

Deep imaging surveys driven primarily by cosmology will, by 2025, provide 0.2 arcsec image quality optical data over 15,000 deg$^2$ from Euclid and multi-band photometry to AB~27 from LSST. Associated low resolution spectroscopy from eBOSS, PFS, DESI and Euclid will characterise the demographics of the galaxy population (luminosities, masses, star formation rates), together with statistical properties of large-scale structure, out to relatively high redshift. The WG estimates that, in addition to massive catalogues of photometric redshifts and well-defined spectral energy distributions, there will also be spectroscopic redshifts for as many as $10^8$ galaxies over 0<z<3. JWST and E-ELT will complement this with higher quality spectra but over much smaller regions of sky.

The obvious challenge for the WG is whether, at that point, there is a strong case for more galaxy spectra. We concluded there are four basic arguments.

1. The majority of these spectra (from Euclid/PFS/DESI) will be low resolution with modest S/N, often with only a single emission line, thus seriously limiting the science that can be done.
2. Deeper spectroscopic surveys will probe relatively small volumes, limiting our ability to trace the cosmic web at high redshift.
3. The spatial density of spectra will be relatively low and biased towards brighter galaxies. This will limit our ability to study the importance of the close environment around galaxies and their influence thereupon. It will also prevent precise measurements of clustering on those small scales relevant for modelling galaxy-halo relations.
4. Finally, the majority of the current and upcoming surveys are optimised for cosmological studies, and are sub-optimal for studying the *physics of galaxy formation on a statistical basis.*

The WG concludes there will be a need for extensive surveys producing large numbers of spectra with sufficient *quality* (good spectral resolution and S/N), over *very large volumes* with *deep exposures* to ensure large volumes sampled, while at the same time providing *high spatial density* information on small scales.

A good example is the Sloan Digital Sky Survey (SDSS), which has provided such high quality spectroscopy for the z < 0.2 Universe. This has been an enormous stimulus to understanding the *physics of the galaxy population* over the last decade, even though the initial motivation was largely concerned with cosmology and large scale structure. The progress achieved with SDSS has inspired attempts to secure equivalent high quality data at intermediate and high redshift (e.g. the DEEP2, VVDS and GAMA surveys). This has been the case even for surveys such as VIPERS whose initial goals, as in SDSS, were primarily cosmological. However, compared to SDSS, these recent surveys have not provided comparable data on large scales due to the limited capabilities of existing telescopes.

So what would a survey facility optimised for galaxy evolution studies look like? Fundamentally, the galaxy formation and evolution process is governed just as much on small scales as on large (cosmological) scales, so a high spatial density sampling of targets is essential. The spectroscopic resolution required extends beyond emission or absorption line redshifts to, for example, resolving emission line doublets and measuring internal stellar velocity dispersions that encode crucial physical information. Integral field spectroscopy may be beneficial providing internal kinematics and population variations. Finally, although a survey optimised for galaxy evolution need not cover the fully accessible sky, it should sample representative volumes implying survey areas >100 deg$^2$ with high density sampling. Collectively, these lead to requirements for very high multiplex gain using optical fibres and/or a massive integral field unit capability

Why has this not been done already? A fundamental reason is that without a careful pre-selection of galaxies in the targeted redshift range, such a survey would be prohibitively expensive due to interlopers. However, by the mid-2020s deep multi-band imaging over very large areas will become available for the first time and, as detailed below, this will permit efficient and reliable target selection. To data, this has only been possible in small areas (e.g. VANDELS in HST survey areas). In what follows, noting the discussion in Section 4.1, we illustrate the remarkable scientific potential of a 10-12m class facility with a 5 deg$^2$ field of view and a multiplex gain of 5000$^2$. With such a facility we can open a new chapter on galaxy evolution studies.

**Scientific motivation – the connection to the cosmic web**

Ultimately we seek to understand how the full distribution function of galaxies in the present-day Universe was assembled, the detailed physical processes involved and how these are connected to and governed by the surrounding dark matter structure. Nearby datasets (e.g. SDSS, 2dFGRS, 2MASS, GALEX and WISE to mention the most prominent) have provided a comprehensive view of galaxies and associated structure in the local Universe (z < 0.2), as a function of stellar mass, stellar content, metallicity, star formation rate, environment, and activity level. GAMA (Driver et al. 2016) has been extending such measures to z~0.3 with a high spectroscopic completeness, while VIPERS (Guzzo et al. 2014) is currently pushing them, with similar statistical accuracy, to z ~ 1 (Figure 6).

---

$^2$ We discuss the competitiveness and flexibility in this parameter choice later in this section.

The crucial next step is to push these studies to z > 1 where the star formation and AGN activity in the Universe was at its peak.

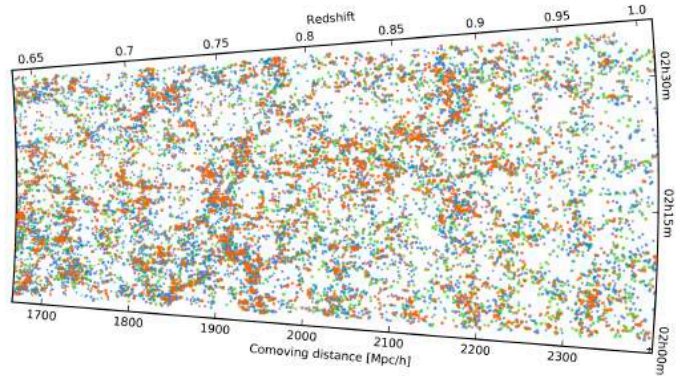

*Figure 6: Galaxy colours in the cosmic web at z=[0.6,1.0], as visible in a section of the W1 field of the VIPERS survey: colours code the U-V colour and the combination of these with the highly-sampled large-scale structure clearly show that the colour-density relation is already in place at z~1. The natural question looking at this picture is when, going further back in time, will galaxies within structures turn blue? (Courtesy of the VIPERS Team).*

The most fundamental feature of hierarchical structure formation is the complex distribution of matter in a structure referred to as the *cosmic web*. Galaxies form and assemble within this evolving network of dark matter halos, filaments and voids, accreting matter from it and ejecting baryons into it. It is therefore the natural focus of galaxy evolution studies and local spectroscopic surveys such as SDSS and 2dFGRS established on a firm statistical basis important relationships between galaxies and their hosting structures. As shown in Figure 6, trends such as the colour-density relation, with red galaxies confined to the high density knots of the cosmic web, are already in place at z~1. Mapping these correlations on similar scales at epochs corresponding to the peak of star formation activity requires deeper exposures over similar or larger volumes. This can only be obtained with a dedicated multi-object spectroscopic facility, optimised selection techniques and novel analysis methods.

The cosmic web presents the backdrop for several important scientific questions that are likely to still be relevant in the 2020s:

- How do the small-, intermediate- and large-scale structures affect the accretion onto, star formation in and ejection of matter out of galaxies?
- How do outflows from galaxies chemically enrich the cosmic web and the intergalactic medium in general? Are there regions of the cosmic web that are chemically pristine at intermediate redshift?

- How do the distribution functions of galaxy properties vary with redshift and environment at z>1?
- What are the properties of the progenitors of the most massive structures found in the present day Universe?

While some of these questions are certainly being addressed by current or near-future surveys, these surveys generally lack the volume necessary for representative results, suffer from poor sampling on small scales and/or fundamentally will not produce spectroscopy of the required quality and resolution. The joint requirements are challenging: volumes of ~ Gpc$^3$ are needed to sample the full range of web environments (e.g. Vikhlinin et al 2009), yet galaxy number densities ~$10^{-2}$ h$^3$ Mpc$^{-3}$ are important to probe filamentary structures and highly non-linear knots. Spectra with high S/N and good resolution yield physical and kinematic parameters for each galaxy that provide the diagnostics required to address the questions above. In present-day surveys, this can only be achieved by the undesirable practice of stacking spectra.

Rather than design such a survey based on existing instruments on present 8-10 m class telescopes (e.g. PFS, MOONS), it is much preferred to consider a facility that can support a flexible survey strategy, supporting multi-wavelength data to facilitate selection and novel analysis methods to push beyond the current boundaries.

**Proposed Galaxy Surveys:**

As an example of the power of starting anew with a wide field 10-12m class facility, we illustrate below a multi-level survey strategy which is designed to exploit targets selected from the dramatic growth in wide-area deep imaging in the coming decade and which would reconstruct the cosmic web to z~4.

**Charting the cosmic web - the million galaxy per Gpc$^3$ survey over 1 < z < 4**

The primary aim of this component is to *reconstruct the 3-D density distribution* with a fidelity comparable to the SDSS in redshift bins of $\Delta z=0.5$ from z=1 to z=4 with $10^6$ galaxies per bin for six Gpc$^3$ volumes. Since the abundance of clusters similar to Coma (halo mass ~$10^{15}$ M$_\odot$) is several per Gpc$^3$, this ensures we probe the full range of densities, halo masses and environments in each redshift bin.

Such volumes must be sufficiently large to contain all representative structures yet probed finely enough to enable reliable detection of proto-clusters and voids with spectra of sufficient S/N and resolution to enable measures of basic physical parameters such as continuum-based star-formation rates and stellar metallicities derived from UV absorption at z > 2 (e.g. Faisst et al 2016). Such unique data will permit correlations of the basic physical properties of each galaxy in the context of the both the large- and small-scale density environment.

A powerful technique for deriving the associated density of *dark matter structures*, but within a more restricted redshift range, involves charting the 3D topology of the *Lyα forest* seen in absorption line spectra of background Lyman-break galaxies (LBGs). Such spectra provide high spatial resolution probes of the density field along the line of sight, instead of the single data point provided by a galaxy redshift. Long considered the realm of 30m-class telescopes (e.g. Steidel et al 2009; Evans et al 2015), recent applications have shown adequate information is feasible with 8-10m-class telescopes (e.g. Lee et al 2014; Stark et al 2015a,b) provided the continuum S/N and spectral resolution (R~1000-1500) are adequate and, for studies at z~2-3, there is reasonable blue sensitivity (Figure 7). MOSAIC on the E-ELT has the potential to complement such an application with finer sampling but not over the required cosmic volumes.

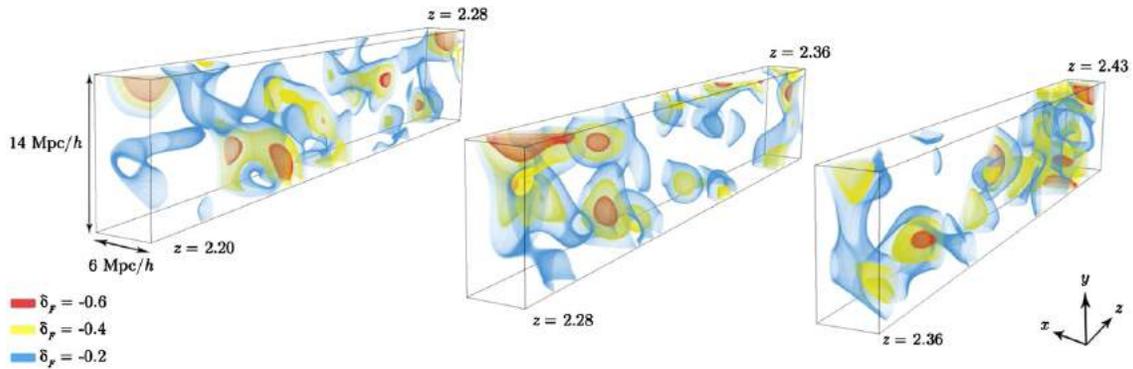

*Figure 7*: Demonstration of a reconstruction of the 3-D dark matter distribution in three redshift slices from Keck absorption line spectra of Lyman break galaxies from the analysis of Lee et al (2014). Colours represent (negative) measures of the transmission which relate to increasingly positive overdensities.

The fluxes and surface densities of the sources required to chart evolution in the cosmic web using both galaxies and the Lyα forest over the common redshift range 2<z<3 are similar - indeed, they are often the same sources. In each Δz=0.5 volume bin, a 1 Gpc$^3$ volume (h=0.7, hereafter) implies a sky area of ~200 deg$^2$. For 10$^6$ galaxies per bin (i.e. a minimum density of ~3. 10$^{-2}$ Mpc$^{-3}$ required to identify filaments and knots), necessitates a target density of ~5000 deg$^{-2}$ bin$^{-1}$ or ~30,000 deg$^{-2}$ across the entire redshift range. To uniformly populate the full redshift range will require careful photometric selection in deep wide field catalogues such as those that will become available with LSST and Euclid. The limiting magnitude will typically vary from $i_{AB}$~23.1 at z~1 to $i_{AB}$~24.9 at z~3. Current 10m surveys spanning this redshift and magnitude range (e.g. Steidel et al. 2003, Stark et al 2009) indicate exposure times of 2-3 hours for emission line redshifts and 5-7 hours for the most challenging i < 25.8 sources to z~4.

In the redshift range 2<z<3, this target surface density is more than adequate to map the cosmic web using the Lyα forest. Lee et al (2014, Figure 7) achieved a spatial resolution of ~3.5$h^{-1}$ Mpc targeting z~2.3-2.8 galaxies with $g_{AB}$~24-24.8 with a surface density of ~1000 deg$^{-2}$. Exposures with Keck LRIS at R~1000 were typically 2 hours achieving a continuum s/n~3 per Å. To push to z~3 with this s/n will require exposures of ~7 hours, comparable to that for the primary galaxy survey.

If the unit exposure for the brightest targets is ~2 hours, with fainter targets studied in multiple visits, we can complete spectroscopic studies at the required surface density (30,000 deg$^{-2}$) over each 5 deg$^2$ field in a sequence of exposures totalling 60-100 hours. Although only an illustrative example, the entire survey of 6 × 10$^6$ galaxies would then take 400 dark nights or so, an achievable goal in say 3-5 years and comparable to the investment being made in similar surveys today (e.g. PFS).

**Astrophysics of galaxy assembly: going deeper over 2 < z < 4**

Galaxies are fed by gas accreted from the circumgalactic medium, while simultaneously return gas via enriched outflows. This *baryonic cycle* is one of the most fundamental aspects of galaxy evolution, yet many key questions cannot be addressed with today's surveys. Does activity in galaxies lead to changes in the covering fraction of the neutral gas surrounding them, what is the dependence of the cycle on the small- and large-scale environment, how do outflow velocities depend on stellar mass? These questions underpin the detailed physics of "feedback".

The stellar population at redshifts z >2 cannot be characterised in detail with existing surveys. This also impairs our understanding of the emission lines seen in rest-frame optical spectra (e.g. Steidel et al 2016; Masters et al 2016). Whereas JWST will make a significant improvement, its spectroscopic samples will remain modest and lack UV coverage where many key features lie.

The dense, lower resolution survey outlined above will have too low resolution and S/N to tackle these questions. To make progress it is necessary to have spectra with R~3000 to be able to disentangle absorption components from the complex circumgalactic medium and to properly resolve stellar absorption lines in the spectra (Jones et al 2013, Figure 8). The spectra will also provide high-S/N measurements of the Lyα emission line. The shape of this line encodes additional information on the circumgalactic medium and, in cases where it is double peaked, it can be used to study the dependence of the escape fraction (Verhamme et al 2015) on stellar mass, star formation rate and environment; important to studies of galaxies at the epoch of reionisation.

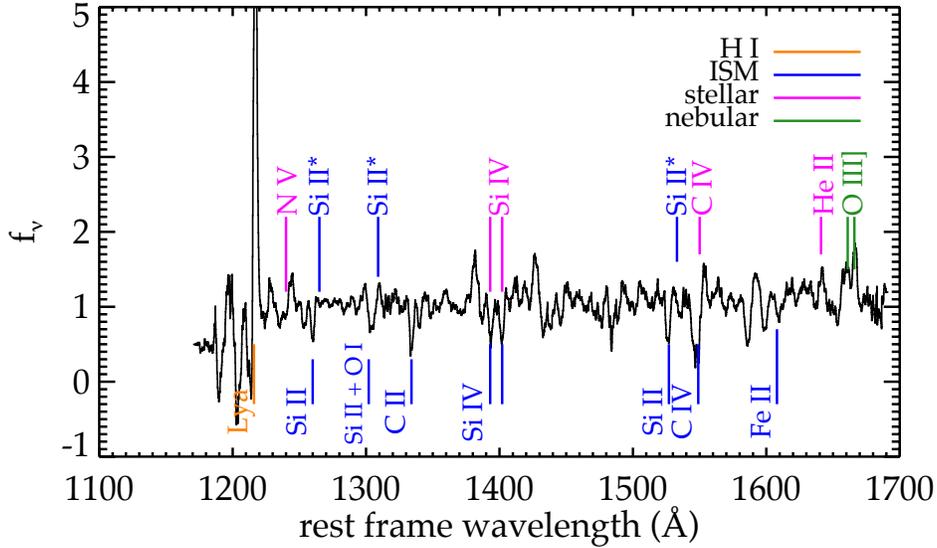

*Figure 8: The merit of higher spectral resolution. A Keck absorption line spectrum at R~3000 for a $R_{AB}$~22.6 galaxy at z=4.1 (Jones et al 2013) revealing distinct diagnostic features from stars, the interstellar gas and nebular emission.*

Such a deeper survey can be conducted with longer integrations on a subset of the galaxies selected for the survey above. A S/N > 10 per Å to allow detailed absorption line spectra of $i_{AB}$ < 24 galaxies can be achieved with exposures of ~20-50 hours. Targeting ~5% of the galaxies in the previous survey would lead to a sample of $2\times10^5$ galaxies with high quality spectra - two orders of magnitude larger than the VANDELS survey. Such a survey would approach the ideal goal of constructing a "SDSS @ z=2-3", especially if combined with MOONS to secure the rest-frame optical spectroscopy.

Given the 60-100 hour total exposure time per position of the cosmic web survey discussed above, and the assumed 5 deg$^2$ field of view and multiplex gain of 5000, these observations can be carried out in parallel by diverting a relevant subset of fibres to R~3000 spectrographs, avoiding any increase in the total survey time.

**The unique power of a wide field and high multiplex gain**

Although the above surveys are illustrative, they demonstrate how, with a large aperture, a wide field and high multiplex gain, ambitious programmes can be completed on a 3-5 year timescale. The WG considers that the total duration of a survey is likely to be an important criterion in making it practical. The quality of spectra necessary to map the cosmic web using the Lyα forest, or examine absorption lines at z~2-4 can never be realised with 4m class facilities. Even in the case of 8-10m facilities such as PFS or MOONS, to chart the proposed 200 deg$^2$ volumes with the required high surface density would be significantly slower (by factors of 3-7) leading to a significantly longer duration survey.

For the surveys discussed, some flexibility in the choice of multiplex gain and field of view may be possible provided the high surface density of fibres is maintained. For example, a slightly smaller field would be acceptable if the multiplex gain was increased.

**An integral view of the IGM – the case for a Super-MUSE:**

To fully understand the interaction between galaxies and the cosmic web, it is necessary to also study the *chemical content of the intergalactic medium* (IGM) via outflows – allowing us to address which type of galaxies dominate its enrichment, the extent to which a galaxy can enrich its surroundings and how this depends on other properties including the overall environment now suitably charted by the above surveys.

The diffuse gas resulting from galactic winds is most readily observed through absorption lines superposed on high resolution spectra of QSOs. It extends to at least one proper Mpc around galaxies at z~2-3 (2 arcmin) in the limited number of cases studied (e.g. Turner et al 2014). For detailed work, it is necessary to associate each absorption system to its responsible galaxy, which can only be achieved via high spatial density spectroscopy of (primarily) emission line galaxies near the QSO sightlines. The ideal instrument is a panoramic integral field spectrograph such as a wide-field successor to MUSE. Such an instrument (here referred to a Super-MUSE) would provide spectra of emission line galaxies down to $m_{AB}$~30 and produce maps of the diffuse gas within their spheres of influence.

Such studies are currently limited to those few quasars for which 8-10m class telescopes can secure high resolution spectra, limiting the statistical power and the ability to study enrichment patterns as a function of galaxy properties. The advent of HIRES on the E-ELT will change this by more than an order of magnitude providing suitable spectra of 3-4 QSOs deg$^{-2}$ to i~20 at z~2-3. Note that the current MUSE is unsuitable for such a program in the redshift range that overlaps with the surveys outlined above, due to its lack of blue wavelength coverage and limited field of view.

A Super-MUSE with a field of view of 3x3 arcmin and blue sensitivity ($\lambda$ < 3600Å) would chart the absorbing galaxies up to 1 Mpc radially around each QSO, mapping the IGM with high quality in ~10-20 hours. There will be 15-20 QSO sightlines within the FoV of the proposed instrument and it may even be possible survey these concurrently with the large-scale fibre surveys outlined above. Over the duration of the survey the enrichment pattern of the IGM will be measured around 800 sightlines, an order of magnitude more than can be done with MUSE on VLT and the Cosmic Web Imager on Keck prior to the advent of the E-ELT.

At the same time, the SuperMUSE data would also provide resolved dynamical and metal abundance maps of ~10-30 galaxies arcmin$^{-2}$ resulting in a final sample of ~10,000-20,000 galaxies at 0.3<z<1.5, an order of magnitude more than what MUSE could deliver by the mid-2020s, and comparable to the impressive low-redshift IFU samples assembled by MaNGA, SAMI and CALIFA.

**Cosmology from the evolution of large-scale structure**

Our discussion thus far has focused on the assembly history and chemical enrichment of galaxies in the context of their interaction with the growth of large-scale structure. The relevant spectroscopic data will, however, also have unique cosmological importance, complementing parallel cosmological surveys which cover larger volumes at lower density with limited information on the properties of the galaxy tracers.

For example, the proposed survey volumes are sufficient for competitive detections of Baryon Acoustic Oscillation and Redshift Space Distortion signals (both from galaxy and Lyα absorber correlations), the primary probes of dark energy and modified gravity that are the focus of projects like Euclid and DESI. However, our proposed facility will provide more representative and physically understood samples of the galaxy population at higher spatial densities. This will widen the information content significantly, allowing new approaches that can exploit *sub-populations* to trace the underlying density field with different bias values. Such techniques are currently in their infancy and include, for example, full Bayesian reconstruction of the statistical properties of the density and velocity fields (e.g. through Wiener filtering, Granett et al 2015), or selecting and combining different galaxy tracers to minimise statistical and systematic errors in redshift space distortions (e.g. McDonald & Seljak 2009). Dense sampling will also enable the characterisation of *cosmic voids* and the tracing of velocity outflows through void-galaxy correlations (c.f. Hamaus et al 2016) offering new probes of both the cosmological background and the growth of structure (and thus of possible modifications of General Relativity).

**Synergies and Requirements**

The successful execution of large-scale z>1 redshift surveys depends on an efficient selection strategy. In recent years this has been demonstrated by surveys such as VIPERS, VUDS and VANDELS which have relied on high quality photometric redshifts obtainable in areas with deep, multi-wavelength photometry such as the HST CANDELS and COSMOS fields. The key point driving the case for our proposed extragalactic surveys is the upcoming bounty of wide-field deep imaging data enabling equivalent target selection over much larger fields. We argue this is a very fundamental synergy – the photometric surveys are limited in their scientific return by the lack of spectroscopic data, and the spectroscopic surveys would be unable to assemble large samples in an efficient manner without the photometric surveys.

Beyond this essential link between the photometric and spectroscopic surveys, the proposed facility will also have strong synergies with near-IR multi-object spectrographs such as MOONS. MOONS will have a smaller field of view but can very efficiently be combined with e.g. the high S/N survey to provide the diagnostic information present in the rest-frame optical.

The facility proposed here would be highly synergistic with the MOS capabilities of the E-ELT (Evans et al 2015), the latter targeting a much smaller field of view. The E-ELT will be an extremely powerful and deep probe of galaxy and structure evolution on fine (~1-10 Mpc) scales, but such programmes will be considerably enhanced by the more detailed environmental context that can only be provided by the larger-scale studies discussed here. We envisage that the E-ELT will study, in exquisite detail, regions of the cosmic web selected from structures charted by the proposed facility.

The Table below summarises the salient aspects of the various science cases discussed above.

| Science case | Sample needed | Wavelength range [Å] | Resolution | S/N per Å |
|---|---|---|---|---|
| **The cosmic web** | | | | |
| Tomographic reconstruction | 500/deg$^2$ | 3600-5000 | ~1000 | 3 |
| Proto-clusters | > Gpc$^3$ | 3600-7000 | ~1000 | 1-2 |
| **Dense, large redshift survey** | | | | |
| Scaling laws and their evolution | >10$^5$ and multiple z-bins | - | - | - |
| Outflows, stellar pops – coarse | >10$^5$ | 3600-9000 | ~1000 | 4-5 |
| z<2 survey, redshifts | - | 4000-1.1µm | ~1000 | 1-2 |
| z>2 survey, redshifts | - | 3600-8000 | ~1000 | 1-2 |
| **Deep, sparse redshift survey** | | | | |
| Outflows – in detail | >few x 10$^4$ | 3600-7400 | ~3000 | 10 |
| Stellar populations | >10$^5$ | 3600-9000 | ~1000 | >10 (detailed) |
| **"Super-MUSE"** | | | | |
| IGM enrichment | >500 sightlines, FoV 3'x3' | 3250-6000 (IFU) | >2000 | 1-2 |

Clearly some versatility in spectral resolutions (1000 < R < 3000) and wavelength range is required. While many absorption line science cases will benefit from blue sensitivity, the z < 2 surveys will also strongly benefit from an extended red/near-infrared sensitivity to allow the detection of [O II] emission. A facility with a high multiplex gain (N~5000) and a wide field (~5 deg$^2$) has been assumed in crafting outline survey strategies. A wide-field IFU, as an additional capability, will enable unique surveys of the galaxy-IGM connection around many hundreds of sightlines to QSOs studied by the E-ELT.

## 4.3 Transient and Other Science

**The context**

Transients are defined here as unique or quasi-unique sources that significantly increase in brightness and fade over a timescale of hours to months. We include variable AGN in this context since regular monitoring covers many of the same practical issues as large scale transient observations. Transient science is one of the major growth areas in astrophysics. The number of transients being found by facilities such as PanSTARRS and PTF is increasing rapidly and major new facilities are being developed to further progress such as the Zwicky Transient Facility (ZTF) at Palomar and, ultimately, LSST. LSST is expected to find more than a million transients in its first five years. Many communities are now considering the spectroscopic requirements for follow-up of LSST sources and this topic forms the basis of this section. The overall transient science case is summarized in detail in available LSST documentation (LSST Science Collaboration 2009) from which we draw relevant numbers and figures.

**Supernovae: cosmology and understanding the phenomena**

Type Ia SNe have been studied intensively over the past two decades because of their role as cosmological distance indicators. Careful analyses of homogeneous samples, e.g. ~500 SNe Ia from the Canada France Hawaii SN Legacy Survey (Conley et al 2011) suggest that a limit has been reached for cosmological applications by systematic uncertainties in photometric calibration and host galaxy dependences. There is therefore widespread belief that other techniques (baryonic acoustic oscillation and weak lensing) will be more effective in the coming decade (Weinberg et al. 2013). Nonetheless, it is important to continue investing in SN Ia science both to make progress in studying this poorly understood phenomenon which will reduce these systematic uncertainties. Moreover, increasing the number which probe the redshift-luminosity relation into the decelerating era beyond redshift z~1 will be particularly effective given this is where the constancy or otherwise of the dark energy density can be tested. Starting in 2022, LSST will increase the number of SNe Ia by 2-3 dex each with well-measured light curves. Follow-up spectroscopy of this vast supply of Ia's is necessary since photometric typing alone leads to significant contamination by other transients (Sako et al 2011) which is particularly problematic for z > 1 samples. Spectra provide accurate redshifts, host galaxy properties and further our understanding of the Ia phenomenon. It is reasonable to also believe it will lead to substantial progress in reducing the current systematic uncertainties.

Core collapse Type II SNe will be about half as numerous as Ia's. Although a case can be made for probing cosmology independently with these events (Nugent et al. 2006), scientific interest has largely focused on understanding the explosion mechanisms and local environments. Associated spectroscopy will provide detailed information on the redshift, progenitor metallicity and stellar mass.

In the spectroscopic follow-up of SNe, it is important to distinguish between *live* and *transpired* events (strictly their host galaxies). As LSST surveys its accessible sky every night, it would take a formidable effort to follow-up *every single event in real time*. However, given reliable photometric typing of SNe, there is a strong case for securing host galaxy spectroscopy, providing accurate redshifts and physical details *after the SNe have faded*. Although LSST photometry will lead to host galaxy photometric redshifts that could be as accurate as $\Delta z \sim 0.02$ (Lidman et al. 2013), several percent uncertainties for the majority of cases, as well as the potential for percent-scale systematic bias in the photo-z determination will be problematic in attempting precision cosmology. As we show below, LSST can readily deliver, over a period of months, a high surface density of transient-related targets, making multi-object spectroscopy with a wide field facility a practical proposition. Indeed, the cases for SN Ia and SN II follow up have similar observing requirements.

*Sky density estimates*

According to the LSST science book (Chapter 8, Table 8.2), the expected rate of transients from LSST is about 300,000 per year (Figure 9), and this is dominated by SNe Ia (2/3rds) and SNe II (1/3rd) with only ~5% in other types of SNe. Of course for events beyond a redshift 1, the predicted rates are quite uncertain, especially for SNe Ia.

Over five years, the total rate of SNe discovered by LSST is expected to be ~80 deg$^{-2}$ making multi-object campaigns practical. The AAT 2dF spectroscopic follow-up of transients found in the Dark Energy Survey, OzDES (Yuan et al. 2015), provides a useful benchmark for our purpose as it combines the spectroscopic follow-up in DES fields of live transients, transient host galaxies and variable AGN. Within the AAT 2dF field, ~250 fibres can be effectively used for transient follow-up, of which ~100 are dedicated to live and faded SNe targets. The projected number of transients from LSST seems consistent with the SN numbers from DES (~*2500* in 5 years over 27 deg$^2$, Lidman et al. 2013) for candidates brighter than r<*23* (Yuan et al. 2015, Figure 10).

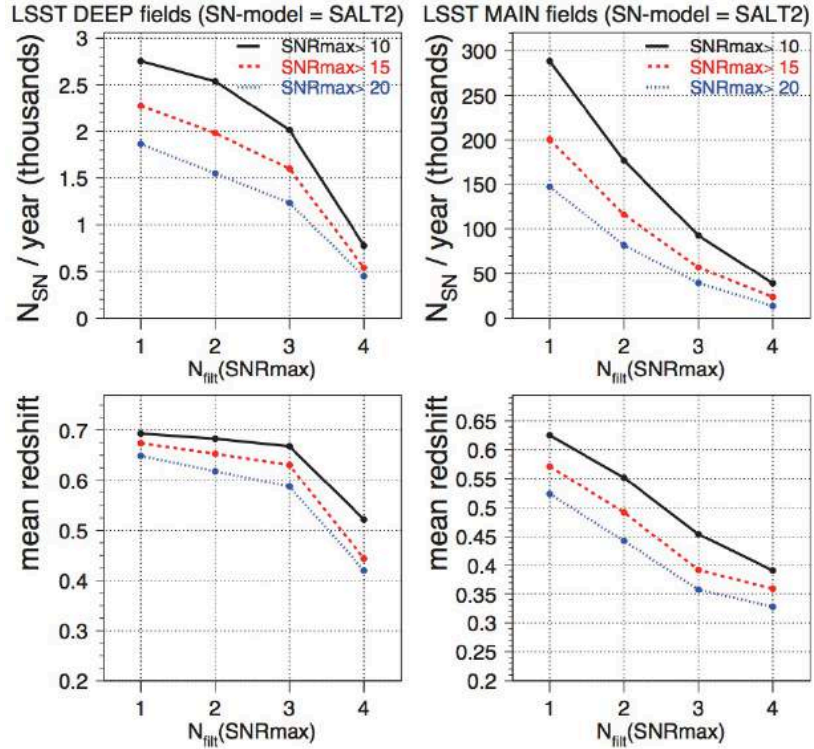

*Figure 9*: Upper panels show the number of SNe Ia detected in a single year in the deep drilling fields (left) and universal cadence fields (right) which have good enough photometry to allow fitting of a high-quality light curve. The more filters in which high-quality photometry is available, the better the resulting constraint on supernova parameters; the number of filters is shown along the x-axis (from http://www.lsst.org/sites/default/files/docs/sciencebook/SB_11.pdf ).

*Live transients*

Assuming a mean observable time of 50 days for a live SN, this provides about 40,000 events visible at once over the southern sky – about 2 deg$^{-2}$ at any instant. OzDES typically include 5–10 active transients with $r$ <23 in the 2dF field, which is consistent (Yuan et al. 2015). The density will be twice as high in the LSST deep-drilling fields. Even for a large field of view (e.g. ~5 deg$^2$), only 10–20 live targets can be studied simultaneously. While still an advantage over current instruments such as X-shooter, clearly it does not drive the case for such a wide-field facility. Such targets would have to be combined with other programmes reaching to a similar depth.

However, assuming a 10-12m class telescope, since ~15 fields to $r$~23 could be observed per night, ~10–20% of all LSST transients could be observed spectroscopically with a merged programme. Over LSST's lifetime, this has the potential of providing a spectroscopic database of 150,000-300,000 SNe enabling a raft of new analyses. Detailed analyses of subclasses would likely significantly reduce systematic biases in distance measures and offer a new route to understanding of how stars explode. The key challenge is therefore how to merge such a *live transient survey* with other galaxy surveys requiring a similar exposure time and spectral resolution.

*Host galaxy follow up*

The likely mean redshift for LSST-discovered SNe Ia is *z=0.57* (Figure 10). 95% of their hosts are brighter than *r<24*. OzDES reaches 50% redshift completeness at *r~22.3*, and Yuan et al. (2015) argue that OzDES will reach 80% completeness for targeted SN host galaxies. Likewise 4MOST is planning to follow-up host galaxies from LSST in a similar manner to z~0.8. Both surveys would use only about a hundred fibres at a time in conjunction with other multi-object targets.

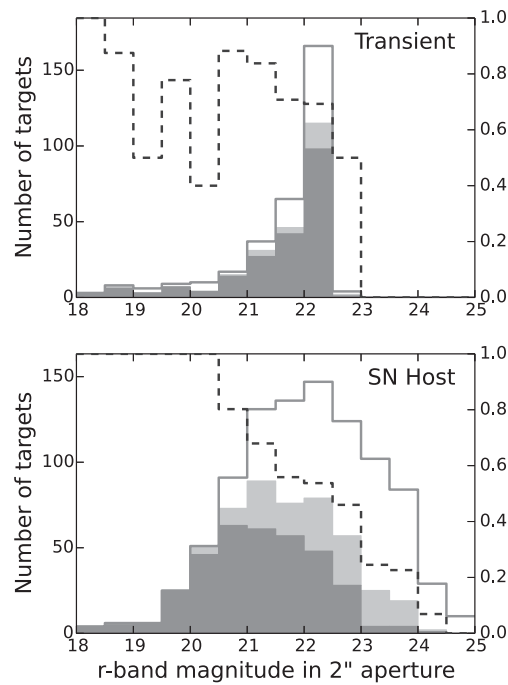

*Figure 10*: *Redshift completeness as functions of r-band magnitude measured in a 2" aperture, for transients (top) and SN host galaxies (bottom). Dark shaded histograms represent objects with the most reliable redshifts. Light shaded histograms represent objects with reliable redshifts. Unfilled histograms are for all targets. Completeness (as dashed curves) is defined as the fraction of objects for which at least reliable redshifts are obtained. A few galaxies in the bottom panel have magnitudes fainter than 25. These bins have low completeness and are excluded to emphasize the more typical magnitude range (from Yuan et al 2015).*

This suggests the main focus of a 10-12m facility should be to targets the hosts of more distant events where the rates are more uncertain. For host galaxy confirmation, optical spectroscopy is likely to be adequate given [O II] can be observed over the redshift range 1<z < 1.6 where the current SNe Ia samples are limited and the interest in dark energy evolution is high.

*Spectroscopically-discovered SNe*

In addition to follow-up of photometrically detected SNe from LSST, a wide-field multi-object facility will also serendipitously find SNe as a byproduct of its own galaxy surveys (Section 4.2). Such a sample would have very different selection methods and a subset would be located very early in the phase, perhaps in the first days after explosion. For a galaxy survey targetting 20,000 galaxies per night, an early SN would be discovered every few nights. These would form the basis of follow-up imaging and spectroscopy with other facilities.

**New discovery space with rarer transients**

Although it is hard to predict transient science discoveries ten years hence, it is certainly conservative to focus entirely on gathering spectra for hundreds of thousands of Type Ia and Type II SNe. In addition to significantly expanding samples of rarer known transients, we can also expect many surprises. In the following section we explore these rarer transients recognising, as with SNe Ia and II, that their surface densities will also be low. While not driving the requirements for a high multiplex gain, a wide field is essential for their efficient follow-up alongside galaxy surveys. So long as such targets can be included other surveys, exciting science will emerge.

*Stripped stars*

The current sample of stripped envelope SNe (Types Ib and Ic) is small. They contain high-velocity Ic's that may be engine-driven explosions and some of which are the SN counterparts to gamma-ray bursts. Major questions surround the explosion mechanism, the mass loss rate and the final stages prior to explosion. While a large number of these will be found by LSST, only limited information can be gleaned from their light curves. Following the survey outlined above that could provide spectroscopy for 10–20% of LSST SNe, this would give 1500–3000 spectra of SNe Ib/Ic. This is about an order of magnitude larger than current samples (e.g. Modjaz et al. 2014) and enough to do detailed population studies.

*SLSNe, other core collapse and tidal disruption events*

Similarly, other known classes of SNe can be spectroscopically observed (e.g. IIn, IIL, etc.). The new classes of SNe or SN-like transients found by large area photometric surveys, such as superluminous supernovae, calcium-rich gap transients, tidal disruption events etc. will also be found in great abundance by

LSST. A wide-field spectroscopic facility will be ready to deliver spectra of these in numbers large enough to strongly impact our understanding of them.

*Gravitational wave counterparts and kilonovae*

Electromagnetic counterparts to gravitational wave sources may be found at a rate as high as 50-100 per year from 2020, with error-regions of a few tens of square degree once a third, sensitive observatory comes online. While black hole – black hole binaries are not expected to offer EM signatures, neutron star mergers should deliver a signal at red or near-infrared wavelengths, depending on the optical depth of the neutron-rich ejecta (the so-called 'kilonova' signature, Tanvir et al. 2013, Barnes et al. 2016). A wide field spectroscopic telescope will aid discovery, providing the deepest spectroscopy of every credible source in the error-region. However, imaging will doubtless remain the primary tool for discovering counterparts.

**Active Galactic Nuclei**

*Motivation*

Spectral variability monitoring of broad line AGN permits reverberation mapping (RM) of the broad-line region (BLR, Bentz et al. 2013). Such RM allows direct estimates of black hole (BH) masses to be made. Currently, supermassive black hole mass (SMBH) estimates are made at low redshifts using gas and stellar dynamics and RM of the H$\beta$ line. At higher redshifts only indirect SMBH mass estimates are available, using the radius-luminosity relation and single epoch spectra (e.g. Vestergaard et al. 2008). These estimates rely on many assumptions and have considerable scatter and systematics (e.g. Denney et al. 2009). RM of broad line AGN could provide direct estimates of black hole masses out to $z$~4, and provide a calibration for single epoch mass estimates out to $z$~7. Broad line RM can also establish the radius-luminosity relation for AGN at $z > 0.3$, a potentially useful cosmological distance indicator (Watson et al. 2011, King et al. 2014). Finally, such data would also help to understand the structure of AGN feeding and the interaction between the BLR and the accretion disk.

*Reverberation-mapping of thousands of AGN*

A wide field large aperture spectroscopic facility could undertake an AGN RM survey of a sample two orders of magnitude larger than the ~60 AGN at present (Bentz et al. 2013, Yue et al. 2016). Current surveys with OzDES (King et al. 2015) and SDSS (Yue et al. 2015) offer a good idea of what is possible. Both are using a multi-object capability for the first time and thus do not have an explicit SNR basic requirement. However, to be competitive for cosmological purposes, a SNR~7 per Å is required, so a sample limited at $r$ ~22 would be straightforward for a 10-12m class telescope. Going fainter is certainly possible dependent on the temporal baseline and source variability. Within a field of 5 deg$^2$, several thousand high quality AGN spectra would be obtained in a night. Those fields would then need to be revisited approximately 50–100 times with a cadence of 5–30 days, a prospect well beyond anything but a dedicated facility. High quality, high-cadence photometric light curves are invaluable in pinning down the spectral lag (King et al. 2016) and would be readily available from LSST. The resulting data for each AGN would surpass in coverage, uniformity and quality, anything so far achieved even locally (well-studied cases like NGC5548 would be comparable). This would dramatically improve the reliability of the current SMBH mass functions, improve our comprehension of the SMBH environment, and possibly provide a new, entirely independent cosmological distance tool, with its own absolute calibration (Hönig et al. 2014).

**Conclusions**

All of the targets discussed above are relatively sparse on the sky compared to Galactic stars and faint galaxies. Even transient hosts and AGN have maximum sky densities about 100 deg$^{-2}$. Clearly a large field of view and collecting area are required to make optimal use of a multi-object spectrograph for these science cases. Even so, the multiplex utility is limited emphasizing the need for coordination with other (most likely) galaxy surveys. Nonetheless, the efficiency of such a facility for transient studies should be considered on the science return, not simply the multiplexing gain. The arrival of large field-of-view, synoptic imaging surveys (currently PTF and DES, soon ZTF and LSST) indicates that our proposed spectroscopic facility can readily take advantage of the supply of fresh transients at moderate sky densities leading to a factor of several hundred increase in numbers of spectra of SNe, exotic transients, early SN spectra, transient hosts, and reverberation-mapped AGN. It is admittedly hard to be precise on how this increase in sample size will impact our study of these sources, but it is likely to be similar to the impact of the SDSS on studies of galaxies.

**Requirements**

The wavelength range for SN and host galaxy studies will depend upon the emphasis given to follow-up distant events. For z > 1 host galaxy follow up the optical window (400-1000nm) will remain adequate but for live z>1 SN studies, access to the region up to 1.3µm would be advantageous. The spectral resolution required for galaxy surveys (R~3000) is quite adequate for typing most transients, for host galaxy redshifts and gas-phase metallicity measures.

For quasar RM, the key is measuring the flux of the broad lines, so even lower resolution (R~300) would be adequate. A broad spectral range would permit cross-calibration of the RM signal for different emission lines. The full optical range (400–1000nm) is sufficient to observe the primary emission lines, from Hβ at z~0 to CIV 155nm out to *z* > 5.

The low occupation of fibres means that their compatibility with other science cases should be examined. The requirement for a very large field of view is similar to that for Galactic archaeology (Section 4.1). However, the stellar spectroscopy case requires high resolution. If low and high resolution spectra can be undertaken simultaneously, transient science can be incorporated into a high Galactic latitude survey. A more straightforward marriage is likely for galaxy surveys for which the instrumental requirements are very similar.

# 5. Summary of Requirements

Each of the 3 science cases described in the previous section discussed its own instrumental requirements but, for convenience, here we summarise and briefly justify the key parameters. We consider it useful to separate discussion of the telescope, its field of view and multiplex gain, from the various spectroscopic instruments, which may be designed and implemented over a longer period. In the following section we discuss the practicality of these choices in the context of preliminary design work undertake in conjunction with ESO staff.

**Telescope parameters:** In both the Galactic and extragalactic cases, a 10-12m aperture is required to respectively secure very high resolution stellar spectra for detailed chemical studies and absorption line work on faint galaxies. Given the likely availability of PFS (and possibly MSE), such an aperture is desirable in the southern hemisphere to be competitive. As was demonstrated with SDSS and the AAT 2dF facility, versatility in accessing/re-instrumenting the focal plane(s) is highly advantageous, for example, in considering the merits of both a massively-multiplexed fibre capability as well as a panoramic Super-MUSE.

**Field of view:** A wide field of ~ 5 deg$^2$ will enable a crucial multiplex advantage to be employed for $m$ < 24-25 galaxies visible to redshifts $z$~4 and for a complete survey of 30-85 million $V$ < 17 stars that can completed on a reasonable timescale. A wide field is also optimal for the rarer transient population and their galaxy hosts. Such a large field, if achievable, would represent a significant advance over PFS (1.25 deg$^2$) and MSE (1.75 deg$^2$).

**Multiplex gain:** The surface density of accessible Galactic stars and m < 25 galaxies demands a high multiplex gain to efficiently exploit the necessary panoramic field of view. For the surveys proposed we adopted a fiducial number of 5000 fibres, about twice that for PFS and MSE but similar to that being proposed for DESI on the 4m Mayall telescope. This will represent a major challenge in terms of the number and cost of the spectrographs required.

**Spectral resolutions:** The WG study has determined the main advance in multi-object spectroscopy lies not simply in the scope of the surveys nor their depth in terms of limiting magnitude, but from the quality of the resulting spectra, particularly for the Galactic case where resolutions R ~ 20,000 to ≥ 40,000 are required for accurate and extensive chemical tagging. This contrasts with the much more modest requirements of the extragalactic case where resolutions of R~1000-3000 appear adequate.

**Wavelength range:** The recommended wavelength range emerging from the WG study is 360 – 1000 nm. Access to the blue spectral region (360-500nm) is required to probe the Lyman alpha forest in the 2.5 < z <3.5 redshift range where galaxy assembly and the influence of the cosmic web is particular important to unravel. For stellar composition studies of a wide range of neutron capture elements, the blue spectral region is also crucial as the relevant absorption features are mostly concentrated below 450nm. Access redward to 1000nm is required for studies of host galaxies of SNe up to redshifts z~1.5.

Access to the near infrared spectral region beyond 1 micron is being offered by PFS (<1.3 microns) and MSE (<1.8 microns) and of course MOONS, but such an extension would come at a considerable extra cost in detectors for the proposed facility. An infrared arm for the extragalactic spectrographs would enable efficient survey capability in the 1.5 < z < 2 `redshift desert' as well as studies of nebular lines over 1.5 < z < 5 yielding systemic velocities of survey galaxies from which outflows can be characterized in conjunction with the local cosmic web measures. It would also permit reverberation mapping studies of QSOs using additional lines such as Hβ to redshifts z~1.7. However, coordination with MOONS may obviate this need and should be considered carefully.

**Panoramic IFU:** The concept of a Super-MUSE follows from the success of MUSE in its first year of full science operations and the WG considered briefly the possible scientific advantages of a wider-field version, noting such an instrument could not feasibly be mounted on the VLT and, possibly, the VLT and E-ELT will focus in the future on AO-fed IFU science. With this in mind, the primary restrictions of the current MUSE are its limited blue sensitivity (> 465nm) and modest spectral resolution (R~2000-3700). For a fundamental advance, the WG considered a Super-MUSE should ideally offer a 3 x 3 arcmin field with resolutions up to R~5000 over the wavelength range extending down to 325 nm. We recognise, however, this may require significant innovative developments.

**A Possible Facility and Its Evolution:** Although the WG was not charged with recommending a specific telescope and instrumentation, it is helpful to sketch one consistent with the above recommendations. We foresee a 10-12m class wide field telescope which can be flexibly used over several decades exploiting several focal stations. It might commence operation with a massively-multiplexed fibre positioner at a Cassegrain focus feeding an array of either low or high resolution optical spectrographs (or even a mixture). Future upgrades could include extensions into the near-infrared spectral region, higher multiplex capabilities and independent instruments such as a panoramic IFU (`SuperMUSE') at a second focal station. The underpinning infrastructure would be the telescope and its unique combination of aperture and field of view.

## 6. Design and Feasibility Issues

In interpreting its charge (Appendix A-1), the WG considered a detailed technical design beyond the scope of the current 6 month exercise. Nonetheless, we distilled several questions relating to feasibility that were briefly addressed at this stage. We discuss these in turn below.

**1. Is it practical to contemplate a 10-12 metre class telescope with a ~5 deg$^2$ field of view with versatile focal planes?**

The WG invited input on this question from Luca Pasquini and Bernard Delabre (ESO); the latter has acted as an optical consultant for the MSE project. Three contrasting telescope designs were considered in conjunction with the WG chair.

(i) A f/3 prime focus design similar to that adopted by Subaru and MSE. The field of view in this case is governed by the size of the prime focus corrector and its atmospheric dispersion compensator (ADC) which is limited by available glassware to less than 2 deg$^2$. The complex fibre positioner must be at prime focus with lengthy fibres feeding spectrographs elsewhere in the observatory building reducing blue sensitivity. In the case of PFS and MSE, such design choices were forced by the existing telescope infrastructure but the WG considered it would not recommend this design starting *ab initio*.

(ii) A novel design published by Pasquini et al (2016) which is based on a 3 mirror anastigmat capable of feeding simultaneously a narrow field super-MUSE and a gravity invariant focal plane with an array of steerable mirrors guiding target light into thousands of IFUs across a 1.8 deg$^2$ field. The merit of this design is dependent on the appropriateness of using deployable IFUs (see below).

(iii) A wide field Cassegrain with a field of view of 5 deg$^2$ at a convenient f/3 focus for fibres, also ensuring a compact telescope and enclosure. The image quality across the field is better than 1 arcsec over 360-1300 nm. This design also has the option of a gravity-invariant remote focal plane with improved image quality suitable for a super-MUSE. The versatile concept is illustrated in Figure 11 and also discussed in Pasquini et al (2016). The WG finds this design particularly appropriate given its recommended parameters in Section 5.

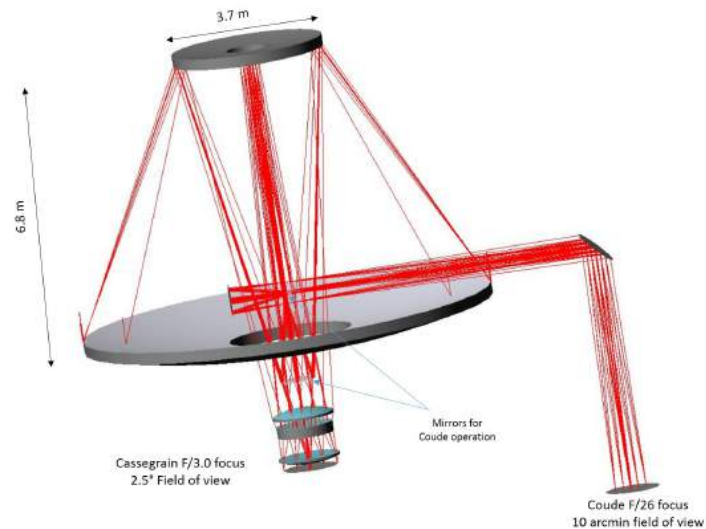

*Figure 11*: *A 10 metre Cassegrain telescope with a corrected field of view of 4.9 deg$^2$ which is limited by the maximum size of available corrector lenses (1.5m). In the configuration shown, two mirrors mounted in front of the corrector activate a 10 arcmin, gravity-invariant, Coudé focus suitable for a super-MUSE (Pasquini et al 2016).*

**2. What role might multiple deployable integral field units play in a next generation wide field spectroscopic facility?**

Given ESO's successful investment in deployable IFUs (dIFUs) for GIRAFFE and KMOS and a novel design (above) for a highly-multiplexed dIFU system for a 10-

12m aperture telescope, the WG explored the scientific utility of dIFUs as part of its study. It would seem that, strategically, they represent the next logical technical advance after multi-fibre surveys.

The advantages of dIFUs over single aperture fibres are both scientific and technical. Scientifically they offer the prospect of resolved galaxy properties (kinematic and chemical) and the ability to deblend close sources. Technically, they ensure no loss of spectral resolution, no aperture losses (or no reliance on an atmospheric dispersion compensator).

However, the WG determined that there are several disadvantages. Typical dIFUs have sizes of 3 × 3 arcsec which, for a given detector real estate, significantly reduces (by ×10) the multiplex gain. Moreover, the resolved galaxy information is of limited use in seeing limited conditions. Figure 12 shows how typical z > 1 (L*) galaxies remain unresolved unless they are surveyed behind adaptive optics. Thus whereas a panoramic IFU (super-MUSE) offers unique discovery space, the WG concluded a highly multiplexed dIFU system is not so attractive.

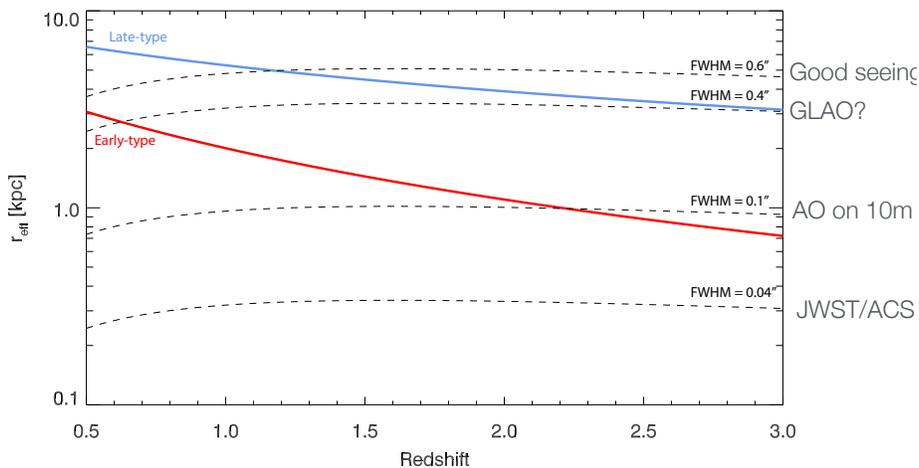

*Figure 12:* *The effective radius of a ~$10^{10.5}$ $M_\odot$ galaxy as a function of redshift for both quiescent and star-forming systems (van der Wel et al 2014). The dashed lines indicate the equivalent scale that corresponds to various image qualities from natural seeing through various degrees of adaptive optics correction on a ground-based 10m telescope and space-based imaging with HST or JWST. For a wide field telescope, only nearby (z < 1.5) galaxies offer useful resolved data.*

### 3. Can adequate sky subtraction be undertaken to satisfy the requirement for high signal to noise absorption line spectroscopy at r~24?

Sky subtraction is an important limitation in spectroscopy and there has always been skepticism in the community about the use of multi-fibre spectrographs as compared to multi-slit instruments which are more generally employed for faint targets. It is primarily for this reason that the WG considers the natural gain in science for a dedicated large aperture spectroscopic facility lies in *improved quality and spectral resolution*, rather than depth. However, understanding the limitations due to systematic uncertainties in sky subtraction is clearly important. The concerns primarily affect the extragalactic and transient applications given the proposed stellar campaigns are primarily targeting V < 17-20 sources. The most challenging requirement is that for the proposed `Astrophysics of Galaxy Assembly' case in Section 4.2, where detailed absorption line studies of 2<z<4 galaxies must be undertaken at R~3000 to $i_{AB}$~24. It is proposed that fainter sources are targeted at lower resolution for the `Million Galaxy per Gpc$^3$' survey, but this programme would rely primarily on emission line detections.

Most sky subtraction techniques currently rely on the use of a fraction of the total number of fibres to estimate the sky contribution. They typically achieve rms level of ~1% of the sky emission via three methods:

*A physical model*: unlike the sky continuum which arises from numerous difficult to model contributions, sky emission lines have a well-known physical origin and occur at specific wavelengths according to precise line ratios. Provided one can model the Line Spread Function (LSF) at each spatial location/fibre (Thatte et al 2012), it is possible to accurately subtract the sky emission.

*Empirical estimates:* The sky emission is estimated from an average of the closest fibres dedicated to sky.

*PCA subtraction:* Since the aforementioned techniques are limited by the accuracy of the physical model or instrumental calibrations (flat fields, scattered light and wavelength calibration), principal component analyses can recover from these limitations by identifying and subtracting appropriate residual signals. The technique has been successfully used by both the BOSS survey (Wild & Hewitt 2005) and in MUSE (Soto et al 2016).

The WG consulted various VLT project groups (FLAMES, SINFONI, MUSE and MOONS) as well as the Subaru/PFS team to examine their methodologies and the measured or expected sky subtraction precision. Hybrid approaches which combine empirical continuum estimates with a physical model of the emission lines have led to sky residual levels as good as 0.6% rms. Accounting for the variations of the LSF across the field of view can lead to significant improvements. Cross-beam switching may provide the ultimate gain whereby each target has an associated secondary fibre. In general, precisions of 1% rms are readily achievable and 0.5% may be possible with extreme care. Simulations of such sky subtraction errors for r~24 galaxy absorption line spectra yielded satisfactory results.

## 4. What are the implications of the wide range of spectroscopic resolutions demanded by the Galactic and extragalactic/transient science cases?

It is not practical to combine the low resolution (R~1000-5000) requirement for the extragalactic and transient applications with the much higher resolution (R~20,000 to ≥ 40,000) capability necessary for Galactic chemical abundance work in a single spectroscopic module. The MSE project has examined this issue in some detail and proposes a low resolution design with two arms (0.36-0.95nm plus 1-1.8 microns) covering the full spectral range, together with a high resolution echelle sampling 3 wavelength slices (360-450nm, 450-600nm and 600-950nm) at R=40,000. Presumably to reduce cost, the multiplex gain is only 1000 for the high resolution capability. This contrasts with 4MOST which has chosen to allocate a fixed fraction of fibres to both high and low dispersion spectrographs. While enabling simultaneous observations, the varying surface densities of Galactic targets is a concern. The mapping of resolution requirements to the various aspects of the Galactic programme is a complex task recommended for a future design phase (Section 8).

In the case of the proposed facility, the low resolution spectroscopic requirement presents no fundamental problem, particularly if a near-infrared capability is abandoned (see Section 5). For the Subaru PFS project with similar requirements, two arms sample the range 380-970nm at R~2300-3000 in a Schmidt collimator plus camera design with each module capable of accommodating 600 fibres. The present instrument requires about twice the investment of PFS, for which each module feeds two 2K × 4K CCD detectors.

Adequate stability is a concern for the high resolution spectrographs and this would preclude a short stretch of fibres as would be optimal for blue sensitivity. Clearly the overall cost of the instrument will be dominated by the high resolution modules, particularly if as required the full multiplex gain of 5000 is assumed.

The WG considered a detailed design for the spectrographic modules will involve likely trade-offs between the key parameters of wavelength coverage, resolution and multiplex gain noting the differing surface densities and expected exposure times for the various survey targets. We recommend this as a key activity in the proposed future conceptual design study.

**Technical Challenges and Innovations**

Despite the above summary regarding the practicality of a large aperture wide-field multi-fibre telescope, the WG recognises there would be many technical challenges to overcome and compromises necessary. There are, however, also opportunities given the long lead time, to perform research into innovative aspects, some of which may reduce the cost and improve performance. We recommend several of these be considered seriously in any subsequent design phase (Section 8) and list here key items worth of serious consideration.

(i) Blue spectral response – this is a key requirement for both the Galactic and extragalactic cases yet optical fibres offer decreasing throughput shortward of 400 nm due to Rayleigh scattering. Short fibre lengths are thus advantageous but may be impractical if the spectrographs are to be on stable platforms.

(ii) The fibre positioners options have evolved significantly over the past 5 years, largely as a result of progress in miniaturisation of electric motors. Substantial performance and cost benefits may be realized by examining new approaches rather than copying proven positioning technologies selected several years ago (e.g. PFS' Cobra design, 4MOST Echidna etc).

(iii) The unit cost of each spectrograph module is likely to be a dominant factor in the total capital requirement (Pasquini et al 2016). This suggests it is worth investing significant effort in both novel designs and suitable (possibly curved) detectors to optimise fast camera designs.

**References**

Pasquini, L. et al 2016 *Proc. S.P.I.E.,* **9906**, ID 03, arXiv:1606.06494
Soto, K.T. et al 2016 *Mon. Not. R. astr. Soc.,* **458**, 3210
Thatte, N. A. et al 2012 *Proc. S.P.I.E.* 8448, ID 09
Wild, V. & Hewett, P.C. 2005 *Mon. Not. R. astr. Soc.*, **358,** 1083

## 7. Operational Considerations

The WG group was asked to review likely operational models for such a facility and the impact on ESO. As a guide we considered the operational models for two relevant facilities.

PFS is a survey instrument on Japan's national Subaru 8 metre telescope funded and constructed by an international team. NAOJ's only technical responsibility is oversight and modifications to the telescope to accommodate PFS. Although the Japanese community will have access to the instrument for small programmes, the primary survey science will be defined via a 5 year SSP (Strategic Science Program) submitted as a single document by the team and assessed via a normal TAC process. Operationally the SSP will then be executed by Subaru staff, with the raw spectroscopic data processed via the team and released into an archive for international use after a proprietary period. No merging of programmes to maximize fibre usage is being considered.

4MOST is also an internationally funded instrument delivered to an ESO telescope. Unlike PFS there is some distinction between the scientific and construction teams with survey operations led by ESO. ESO's technical role is limited to telescope modifications and detector provision. The ESO community will have access to 4MOST via `participating surveys' where targets with similar spectroscopic requirements (exposure time, resolution etc) to those planned by the 4MOST science team can be pooled (by the 4MOST team) to maximise fibre usage. ESO will oversee the balance of team versus community allocations in terms of `fibre hours'.

In the case of the proposed facility, there are both similarities and differences with the above examples. Given the level of resources required for this facility, it is possible that both the telescope and the primary spectroscopic instruments would be funded together, in which case there might not be an external team requesting GTO allocations. If the cost of the spectrographs equals that of the telescope, as seems reasonable (Pasquini et al 2016), it stretches the traditional ESO instrumentation model. However, for such an unique telescope platform which will conduct world-class science for decades, there will also be opportunities for specific novel instruments supplied by individual groups.

Moreover, some of the likely surveys, for example a high spectral resolution southern sky survey of 85 million V<17 stars conducted over a decade, while ambitious, are nonetheless conceptually straightforward. In many ways they are similar to the historical photographic sky surveys undertaken traditionally by national or international observatories on behalf of the community with an additional responsibility for the data products.

However, community involvement is essential in defining, proposing and exploiting these surveys and, for several examples discussed here, the merging of programmes that do not fully occupy all fibres will significantly increase the scientific output. The goal should be to construct a simple model for requesting, reviewing and merging successful proposals alongside the more ambitious multi-year main surveys. A final issue is the responsibility for the data pipeline and archiving. If there is no pre-defined external team, the challenge will be to harness experienced scientists who are keen to maximize the performance of the instrument.

Ultimately, the operational model comes down to (i) defining the scientific programme, (ii) executing the observations and (iii) releasing the data products. For current survey telescopes, these responsibilities are shared between ESO and a distributed team with a clear scientific leadership. The WG considers there are two basic options. ESO could take on more responsibilities (observations and data products) notwithstanding the limited lifetime of the surveys. Alternatively, the community could raise the sums, perhaps through EU initiatives, to manage the entire effort. If the facility is internationally-funded like ALMA, the former is probably a more feasible option.

## 8. Recommendations and International Context

The WG has investigated a limited selection of highly important science programmes that it believes justify serious consideration of a dedicated wide field 10-12 meter facility in the southern hemisphere. Recognising the construction of LSST, the WG also has been in contact with both the MSE team and equivalent US studies (Dodelson et al 2016, Najita et al, to appear) which present a strong case for a similar facility. We believe ESO can maintain a leadership role in this key area of survey science by developing a more detailed proposal.

We thus recommend the programme moves forward over the next year to complete a more rigorous conceptual design study whose goal includes production of a broader and more detailed scientific case, a credible technical design, a cost and schedule. This study would involve two interlocking aspects.

1. The more detailed science case and requirements should be derived via wider community involvement arranged through study groups in several key science areas coordinated by an enlarged WG. The present WG has perforce, during its brief 6 month study, focused on a selection of ambitious surveys in its own area of expertise.

2. Technical effort should be harnessed, both within ESO and in the partner community, to examine (i) the telescope optical design, the accessibility and versatility of its focal planes, (ii) the spectroscopic options (particularly the high resolution requirements) and detector needs, (iii) the fibre positioner and multiplexing arrangements and (iv) operational models and manpower requirements.

To facilitate community participation it is desirable much of these technical studies be undertaken in ESO partner institutions, but managed by a limited core at ESO.

Our final recommendation is that ESO should be open to considering international partnerships in all of the above. A facility of this scope could be considered an optical parallel to ALMA or the SKA. The working group recognises its report arrives at a time of great financial stringency within the world-wide ground-based astronomical community given the preoccupation with realizing ELTs. Nonetheless, by taking this next step and being open to the possibility of a broader partnership with other communities, ESO can maintain a leadership role in what we consider are exciting scientific opportunities, while simultaneously harnessing global interest to realize such a facility.

# Appendices

## A-1 Working Group Details

**Membership:**

Richard Ellis (Senior Visiting Scientist, ESO, Chair)
Joss Bland-Hawthorn (Director, Sydney Institute for Astronomy, Australia)[3]
Malcolm Bremer (Professor, University of Bristol, UK)
Jarle Brinchmann (Associate Professor, Universiteit Leiden, Netherlands)
Luigi Guzzo (Professor, Universita' Statale di Milano, Italy)
Johan Richard (Assistant Astronomer, CRAL, Observatoire de Lyon, France)
Hans-Walter Rix (Director, Max Planck Institut für Astronomie, Heidelberg, Germany)
Eline Tolstoy (Professor, Kapteyn Astronomical Institute, University of Groningen, Netherlands)
Darach Watson (Associate Professor, DARK Cosmology Centre, University of Copenhagen, Denmark)

**Charge to the Working Group**

Following the prioritisation of ESO's programme for the 2020s and the envisaged importance of highly-multiplexed spectroscopy for numerous scientific applications in the future, the ESO Director for Science is establishing a Working Group to investigate the scientific case, synergistic opportunities and practical capabilities of a dedicated ground-based wide field spectroscopic survey telescope in the 2020s.

The Working Group will consider:

- The scientific case for highly-multiplexed ground-based spectroscopy in the era of LSST, E-ELT and space missions such as JWST, Euclid, PLATO and WFIRST/AFTA
- Examples of survey projects and specific synergies with the above facilities and how these affect the design parameters of the survey telescope
- The likely scientific outcomes from the current suite of 4 to 8 metre spectroscopic survey facilities (4MOST/WEAVE, MOONS, DESI/Subaru PFS) and how these propel the case for a more ambitious or complementary facility
- The unique scientific opportunities enabled by a wide-field integral field capability (a la MUSE)

---
[3] Appointed after the first WG meeting

- The mode of operations, data processing and other requirements that relate to disseminating the processed data to the ESO community.

The working group will meet face-to-face and via telecons as appropriate and deliver a report to the Director for Science with its findings and recommendations.

**Schedule**

1. February 8-9 2016  First meeting held at ESO, Garching (full attendance[4])

    Review of charge, discussions with Director-General
    Review of available documents
    Assignments:
        draft science cases, figures of merit for competing facilities
        contacts for external input on science requirements
        review key technical questions
    Discussion of Australian member of WG

2. March 11 2016 Second meeting held at ESO, Garching (full attendance)

    Consideration of draft science cases, figure of merit analyses
    Requirements from space facilities and LSST
    Discussion of operational requirements
    Review of wide field designs (Delabre & Pasquini)
    Assignments:
        3 teams assembled to complete science cases
        Technical notes (sky subtraction, IFUs, future technologies)
        Telescope requirements and further design options

3. April 12 2016 Telecon (full attendance)

    Review of draft presentation to STC
    Delegated sections for final report

4. April 27 2016 Presentation to STC by Ellis – interim report and progress

5. July 31 2016 Delivery of sections of the report to Ellis

6. August 1-12 2016 Consolidation of report and iteration with WG

6. August 23, 24, September 2 Telecons to finalise report

7. September 6 Submission of report to Director of Science

---

[4] Except Professor Bland-Hawthorn appointed prior to second WG meeting

# A-2 Associated Documentation

## Publications

1. *Scientific Priorities at ESO Report*[5] (April 2015)

2. *Shaping ESO 2020+ Together: Feedback from the Community Poll*, ESO Messenger, **161**, 6 (September 2015)

3. *The Detailed Science Case for the Mauna Kea Spectroscopic Explorer*, McConnachie, A. W. et al, arXiv: 1606.00043

4. *A Concise View of the Mauna Kea Spectroscopic Explorer*, McConnachie, A.W. et al, arXiv: 1606.00060

5. *Optimizing the US OIR System in the LSST Era* (National Academy of Sciences, April 2015)

6. *The Science Case for Multi-Object Spectroscopy on the European ELT*, Evans, C. et al, arXiv: 1501.04726

7. *Extragalactic science, cosmology, and Galactic archaeology with the Subaru Prime Focus Spectrograph*, Takada, M. et al, PASJ, **66**, 1 (2014)

8. *Cosmology with the SPHEREX All-Sky Spectral Survey*, Doré, O. et al, arXiv 1412.4872

9. *Cosmic Visions Dark Energy: Science (White Paper for the US Department of Energy)*, Dodelson, S. et al, arXiv 1604.07626

## Conference Presentations (all powerpoints available)

1. *Spectroscopy in the Era of LSST (NOAO 2013)*

2. *ES0 in the 2020s (Garching, January 2015)*

3. *Multi-object Spectroscopy in Next Decade (La Palma March 2015)*

4. *Science with MOS: Towards E-ELT era (Cefalu, September 2015)*

---

[5] http://www.eso.org/public/about-eso/committees/stc/stc-85th/public/STC-551_Science_Priorities_at_ESO_85th_STC_Mtg_Public.pdf

## A-3 Comparing Spectroscopic Survey Facilities

A full list of multi-object spectroscopic facilities and their key parameters can be found in the MSE Scientific Case (McConnachie, A. W. et al, arXiv: 1606.00043) together with their suggested `figure of merit'

$$\eta = A\, \Omega\, N_{MOS}\, f\, /IQ^2$$

where A is the area of the primary mirror, $\Omega$ the instantaneous field of view, $N_{MOS}$ the multiplex gain, f the fractional time available for scheduling and IQ the typical image quality (FWHM). As McConnachie et al admit, it is extremely difficult to encapsulate the scientific performance of any facility with a signal number and, overall, this WG found the concept of ranking facilities by a figure of merit quite misleading.

For example, the WG experimented with a simpler version of the MSE figure

$$\alpha = A\, \Omega\, N_{MOS}\, f$$

recognising most world-class facilities will be on a similar sites or used for challenging work in similar conditions. We considered this number might approach a measure of the *spectroscopic survey speed* for extragalactic applications.

We also considered a figure of merit intended to represent the *spectral information rate* (appropriate for high resolution stellar spectroscopy)

$$\beta = A\, \Omega\, N_{MOS}\, R_{MAX}\, f$$

where $R_{MAX}$ is the highest spectral dispersion available and f the fraction of time devoted to its use.

For completeness we list these figures of merit for what we consider the key facilities in the coming decade, including our proposed 12m spectroscopic survey telescope (Figure 11), dubbed `SpecTel' in the Table below. The two entries for the multiplex gains and schedulable fractions reflect those for low and high dispersion science respectively.

Although, not surprisingly, our proposed facility compares well with present and upcoming facilities, a more detailed inter-comparison reveals the limitations of this kind of exercise. The Table suggests, for example, that Weave is as powerful a facility as Subaru's PFS for both Galactic and extragalactic surveys since they have similar metrics for α and β. In practice, however, the gain in depth or redshift attainable with a larger aperture is not taken into account. Similarly, the gain of a facility than can afford a higher spectral resolution is likely to be non-linear. Although it is convenient to summarise facilities in a single table, each adopts scientific territory appropriate to its unique capabilities and thus comparative measures are extremely hard to construct.

| Facility | $A(m^2)$ | $\Omega(deg^2)$ | $N_{MOS}$ | $R_{MAX}$ | f | log α | log β |
|---|---|---|---|---|---|---|---|
| AAT/2dF | 11.95 | 3.14 | 392 | 28000 | 0.5/0.1 | 3.87 | 7.12 |
| WHT/Weave | 13.85 | 3.14 | 1000 | 20000 | 0.7/0.35 | 4.48 | 7.99 |
| VISTA/4MOST | 13.20 | 4.00 | 1500/800 | 18000 | 0.7/0.3? | 4.74 | 8.35 |
| Mayall/DESI | 12.57 | 7.5 | 5000 | 5500 | 0.5/? | 5.37 | ? |
| VLT/MOONS | 52.81 | 0.14 | 1000 | 20000 | 0.3/0.15 | 3.35 | 8.20 |
| Subaru/PFS | 52.81 | 1.1 | 2400 | 5000 | 0.25/0.1 | 4.54 | 7.80 |
| CFHT/MSE | 99.40 | 1.5 | 3200/1000 | 20000 | 0.5/0.5 | 5.38 | 9.17 |
| SpecTel | 113.10 | 4.9 | 5000 | 40000 | 0.5/0.5 | 6.14 | 10.74 |
| E-ELT/MOS | 1194.6 | 0.011 | 200 | ? | 0.25/? | 2.82 | ? |